\RequirePackage{amsthm}
\documentclass[sn-mathphys,Numbered,pdflatex]{sn-jnl}

\usepackage{graphicx}%
\usepackage{multirow}%
\usepackage{amsmath,amssymb,amsfonts}%
\usepackage{amsthm}%
\usepackage{mathrsfs}%
\usepackage[title]{appendix}%
\usepackage{xcolor}%
\usepackage{textcomp}%
\usepackage{manyfoot}%
\usepackage{booktabs}%
\usepackage{algorithm}%
\usepackage{algorithmicx}%
\usepackage{algpseudocode}%
\usepackage{listings}%
\theoremstyle{thmstyleone}%
\theoremstyle{thmstyletwo}%

\theoremstyle{thmstylethree}%

\begin{document}

\title[]{JWST detection of heavy neutron capture elements in a compact object merger}


\author*[1,2]{\fnm{Andrew} \sur{Levan}}\email{a.levan@astro.ru.nl}

\author[3]{\fnm{Benjamin P.} \sur{Gompertz}}

\author[4,5]{\fnm{Om Sharan} \sur{Salafia}}

\author[6,7,8]{\fnm{Mattia} \sur{Bulla}}

\author[9]{\fnm{Eric} \sur{Burns}}

\author[10,11]{\fnm{Kenta} \sur{Hotokezaka}}

\author[12,13]{\fnm{Luca} \sur{Izzo}}

\author[14,15]{\fnm{Gavin P.} \sur{Lamb}}

\author[1,16,17]{\fnm{Daniele B.} \sur{Malesani}}

\author[3]{\fnm{Samantha R.} \sur{Oates}}

\author[1,4]{\fnm{Maria Edvige} \sur{Ravasio}}

\author[18]{\fnm{Alicia} \sur{Rouco Escorial}}

\author[19]{\fnm{Benjamin} \sur{Schneider}}

\author[20,21]{\fnm{Nikhil} \sur{Sarin}}

\author[21]{\fnm{Steve} \sur{Schulze}}

\author[15]{\fnm{Nial R.} \sur{Tanvir}}

\author[2]{\fnm{Kendall} \sur{Ackley}}

\author[22]{\fnm{Gemma} \sur{Anderson}}

\author[16,17]{\fnm{Gabriel B.} \sur{Brammer}}

\author[16,17]{\fnm{Lise} \sur{Christensen}}

\author[23,24]{\fnm{Vikram S.} \sur{Dhillon}}

\author[15]{\fnm{Phil A.} \sur{Evans}}

\author[19,25]{\fnm{Michael} \sur{Fausnaugh}}

\author[26,27]{\fnm{Wen-fai} \sur{Fong}}

\author[28]{\fnm{Andrew S.} \sur{Fruchter}}

\author[29,30,31,32]{\fnm{Chris} \sur{Fryer}}

\author[16,17]{\fnm{Johan P. U.} \sur{Fynbo}}

\author[1]{\fnm{Nicola} \sur{Gaspari}}

\author[16,17]{\fnm{Kasper E.} \sur{Heintz}}

\author[12]{\fnm{Jens} \sur{Hjorth}}

\author[33]{\fnm{Jamie A.} \sur{Kennea}}

\author[34,35]{\fnm{Mark R.} \sur{Kennedy}}

\author[1,36]{\fnm{Tanmoy} \sur{Laskar}}

\author[37]{\fnm{Giorgos} \sur{Leloudas}}

\author[38,39]{\fnm{Ilya} \sur{Mandel}}

\author[40]{\fnm{Antonio} \sur{Martin-Carrillo}}

\author[41,42]{\fnm{Brian D.} \sur{Metzger}}

\author[43]{\fnm{Matt} \sur{Nicholl}}

\author[26,27]{\fnm{Anya} \sur{Nugent}}

\author[44]{\fnm{Jesse T.} \sur{Palmerio}}

\author[45]{\fnm{Giovanna} \sur{Pugliese}}

\author[26,27]{\fnm{Jillian} \sur{Rastinejad}}

\author[46]{\fnm{Lauren} \sur{Rhodes}}

\author[47]{\fnm{Andrea} \sur{Rossi}}

\author[43,46]{\fnm{Stephen J.} \sur{Smartt}}

\author[46,48]{\fnm{Heloise F.} \sur{Stevance}}

\author[49]{\fnm{Aaron} \sur{Tohuvavohu}}

\author[32]{\fnm{Alexander} \sur{van der Horst}}

\author[44]{\fnm{Susanna D.} \sur{Vergani}}

\author[16,17]{\fnm{Darach} \sur{Watson}}

\author[50]{\fnm{Thomas} \sur{Barclay}}

\author[28]{\fnm{Kornpob} \sur{Bhirombhakdi}}

\author[51]{\fnm{Elm\'{e}} \sur{Breedt}}

\author[52]{\fnm{Alice A.} \sur{Breeveld}}

\author[23]{\fnm{Alexander J.} \sur{Brown}}

\author[4]{\fnm{Sergio} \sur{Campana}}

\author[1]{\fnm{Ashley A.} \sur{Chrimes}}

\author[4]{\fnm{Paolo} \sur{D'Avanzo}}

\author[53,54]{\fnm{Valerio} \sur{D'Elia}}

\author[55]{\fnm{Massimiliano} \sur{De Pasquale}}

\author[23]{\fnm{Martin J.} \sur{Dyer}}

\author[38,39]{\fnm{Duncan K.} \sur{Galloway}}

\author[23]{\fnm{James A.} \sur{Garbutt}}

\author[56]{\fnm{Matthew J.} \sur{Green}}

\author[57]{\fnm{Dieter H.} \sur{Hartmann}}

\author[58]{\fnm{P\'all} \sur{Jakobsson}}

\author[23]{\fnm{Paul} \sur{Kerry}}

\author[12]{\fnm{Danial} \sur{Langeroodi}}

\author[59,60,39]{\fnm{James K.} \sur{Leung}}

\author[23]{\fnm{Stuart P.} \sur{Littlefair}}

\author[2,61]{\fnm{James} \sur{Munday}}

\author[15]{\fnm{Paul} \sur{O'Brien}}

\author[23]{\fnm{Steven G.} \sur{Parsons}}

\author[2]{\fnm{Ingrid} \sur{Pelisoli}}

\author[44]{\fnm{Andrea} \sur{Saccardi}}

\author[23]{\fnm{David I.} \sur{Sahman}}

\author[62]{\fnm{Ruben} \sur{Salvaterra}}

\author[4]{\fnm{Boris} \sur{Sbarufatti}}

\author[2,39]{\fnm{Danny} \sur{Steeghs}}

\author[4]{\fnm{Gianpiero} \sur{Tagliaferri}}

\author[63]{\fnm{Christina C.} \sur{Th\"one}}

\author[64]{\fnm{Antonio} \sur{de Ugarte Postigo}}

\author[65]{\fnm{David Alexander} \sur{Kann}}


\affil[1]{\orgdiv{Department of Astrophysics/IMAPP}, \orgname{Radboud University}, \orgaddress{\street{6525 AJ Nijmegen}, \country{The Netherlands}}}

\affil[2]{\orgdiv{Department of Physics}, \orgname{University of Warwick}, \orgaddress{\street{Coventry, CV4 7AL}, \country{UK}}}

\affil[3]{\orgdiv{Institute for Gravitational Wave Astronomy and School of Physics and Astronomy}, \orgname{University of Birmingham}, \orgaddress{\city{Birmingham}, \postcode{B15 2TT}, \country{UK}}}

\affil[4]{INAF - Osservatorio Astronomico di Brera, Via E.\ Bianchi 46, I-23807, Merate (LC), Italy}

\affil[5]{INFN - Sezione di Milano-Bicocca, Piazza della Scienza 2, I-20146, Milano (MI), Italy}

\affil[6]{Department of Physics and Earth Science, University of Ferrara, via Saragat 1, I-44122 Ferrara, Italy}

\affil[7]{INFN, Sezione di Ferrara, via Saragat 1, I-44122 Ferrara, Italy}

\affil[8]{INAF, Osservatorio Astronomico d’Abruzzo, via Mentore Maggini snc, 64100 Teramo, Italy}

\affil[9]{Department of Physics \& Astronomy, Louisiana State University, Baton Rouge, LA 70803, USA}

\affil[10]{Research Center for the Early Universe, Graduate School of Science, The University of Tokyo, Bunkyo, Tokyo 113-0033, Japan}

\affil[11]{Kavli IPMU (WPI), UTIAS, The University of Tokyo, Kashiwa, Chiba 277-8583, Japan}

\affil[12]{DARK, Niels Bohr Institute, University of Copenhagen, Jagtvej 128, 2200 Copenhagen N, Denmark}

\affil[13]{INAF-Osservatorio Astronomico di Capodimonte, Salita Moiariello 16, 80131, Napoli, Italy}

\affil[14]{\orgdiv{Astrophysics Research Institute}, \orgname{Liverpool John Moores University}, \orgaddress{\street{IC2 Liverpool Science Park}, \city{Liverpool}, \postcode{L3 5RF}, \state{Liverpool}, \country{UK}}}

\affil[15] {School of Physics \& Astronomy, University of Leicester, University Road, Leicester, LE1 7RH, UK}

\affil[16]{Cosmic Dawn Center (DAWN), Denmark}

\affil[17]{Niels Bohr Institute, University of Copenhagen, Jagtvej 128, 2200 Copenhagen N, Denmark}

\affil[18]{European Space Agency (ESA), European Space Astronomy Centre (ESAC), Camino Bajo del Castillo s/n, 28692 Villanueva de la Cañada, Madrid, Spain}

\affil[19]{Kavli Institute for Astrophysics and Space Research, Massachusetts Institute of Technology, 77 Massachusetts Ave, Cambridge, MA 02139, USA}

\affil[20]{Nordita,  Stockholm University and KTH Royal Institute of Technology
Hannes Alfvéns väg 12, SE-106 91 Stockholm, Sweden}

\affil[21]{The Oskar Klein Centre, Department of Physics, Stockholm University, AlbaNova, SE-106 91 Stockholm, Sweden}

\affil[22]{International Centre for Radio Astronomy Research, Curtin University, GPO Box U1987, Perth, WA 6845, Australia}

\affil[23]{Department of Physics and Astronomy, University of
Sheffield, Sheffield, S3 7RH, United Kingdom}

\affil[24]{Instituto de Astrof\'{i}sica de Canarias, E-38205 La
Laguna, Tenerife, Spain}

\affil[25]{Department of Physics \& Astronomy, Texas Tech University, Lubbock TX, 79410-1051, USA}

\affil[26]{\orgdiv{Center for Interdisciplinary Exploration and Research in Astrophysics}, \orgname{Northwestern University}, \orgaddress{\street{1800 Sherman Ave.}, \city{Evanston}, \postcode{60208}, \state{IL}, \country{USA}}}

\affil[27]{\orgdiv{Department of Physics and Astronomy}, \orgname{Northwestern University}, \orgaddress{\street{2145 Sheridan Road}, \city{Evanston}, \postcode{60208-3112}, \state{IL}, \country{USA}}}

\affil[28] {Space Telescope Science Institute, 3700 San Martin Drive, Baltimore, MD 21218}

\affil[29]{Center for Theoretical Astrophysics, Los Alamos National Laboratory, Los Alamos,
NM 87545}

\affil[30]{Department of Astronomy, The University of Arizona, Tucson,
AZ 85721}

\affil[31]{Department of Physics and Astronomy, The University of
New Mexico, Albuquerque, NM 87131}

\affil[32]{Department of Physics, The George Washington University, Washington, DC 20052}

\affil[33]{Department of Astronomy and Astrophysics, The Pennsylvania State University, 525 Davey Lab, University Park, PA 16802, USA}

\affil[34]{School of Physics, Kane Building, University College Cork, Cork, Ireland}

\affil[35]{Jodrell Bank Centre for Astrophysics, Department of Physics and Astronomy, The University of Manchester, M13 9PL, UK}

\affil[36]{Department of Physics \& Astronomy, University of Utah, Salt Lake City, UT 84112, USA}

\affil[37]{DTU Space, National Space Institute, Technical University of Denmark, Elektrovej 327, 2800 Kgs. Lyngby, Denmark}

\affil[38]{School of Physics and Astronomy, Monash University, Clayton, Victoria 3800, Australia}

\affil[39]{ARC Center of Excellence for Gravitational Wave Discovery -- OzGrav}

\affil[40]{School of Physics and Centre for Space Research, University College Dublin, Belfield, Dublin 4, Ireland}

\affil[41]{Department of Physics and Columbia Astrophysics Laboratory, Columbia University, New York, NY 10027, USA}

\affil[42]{Center for Computational Astrophysics, Flatiron Institute, 162 5th Ave, New York, NY 10010, USA}

\affil[43]{Astrophysics Research Centre, School of Mathematics and Physics, Queens University Belfast, Belfast BT7 1NN, UK}

\affil[44]{GEPI, Observatoire de Paris, Université PSL, CNRS, 5 Place Jule Janssen, 92190 Meudon, France}

\affil[45]{Astronomical Institute Anton Pannekoek, University of Amsterdam, 1090 GE Amsterdam, The Netherlands}

\affil[46]{Department of Physics, University of Oxford, Keble Road, Oxford, OX1 3RH, UK}

\affil[47]{INAF-Osservatorio di Astrofisica e Scienza dello Spazio, Via Piero Gobetti 93/3, 40129 Bologna, Italy}

\affil[48]{Department of Physics, The University of Auckland, Private Bag 92019, Auckland, New Zealand}

\affil[49]{Department of Astronomy \& Astrophysics, University of Toronto, Toronto, ON M5S 3H4}

\affil[50]{NASA Goddard Space Flight Center, 8800 Greenbelt Road, Greenbelt, MD 20771, USA}

\affil[51]{Institute of Astronomy, University of Cambridge,
Madingley Road, Cambridge CB3 0HA, UK}

\affil[52]{University College London, Mullard Space Science Laboratory, Holmbury St. Mary, Dorking, RH5 6NT, UK}

\affil[53]{ASI/SSDC, Via del Politecnico SNC, I-00133, Rome, Italy}

\affil[54]{INAF/OAR, Via Frascati 33, I-00040, Monteporzio Catone, Rome,  Italy}

\affil[55]{University of Messina, Polo Papardo, Department
of Mathematics, Physics, Informatics and Earth Sciences,
via F.S. D'Alcontres 31, 98166 Messina, Italy}

\affil[56]{School of Physics and Astronomy, Tel-Aviv University,
Tel-Aviv 6997801, Israel}

\affil[57]{Department of Physics and Astronomy \& Clemson University, Clemson, SC 29634-0978}

\affil[58]{Centre for Astrophysics and Cosmology, Science Institute, University of Iceland, Dunhagi 5, 107 Reykjavik, Iceland}

\affil[59]{Sydney Institute for Astronomy, School of Physics, The University of Sydney, NSW 2006, Australia}

\affil[60]{CSIRO Space and Astronomy, PO Box 76, Epping, NSW 1710, Australia}

\affil[61]{Isaac Newton Group of Telescopes, Apartado de Correos
368, E-38700 Santa Cruz de La Palma, Spain}

\affil[62]{INAF/IASF-MI, Via Alfonso Corti 12, I-20133, Milano, Italy}

\affil[63]{Astronomical Institute of the Czech Academy of Sciences, Fri\v cova 298, 251 65 Ond\v rejov, Czech Republic}

\affil[64]{Artemis, Observatoire de la C\^ote d'Azur, Universit\'e Côte d'Azur, Boulevard de l'Observatoire, F-06304 Nice, France}

\affil[65]{Hessian Research Cluster ELEMENTS, Giersch Science Center, Max-von-Laue-Stra\ss e 12, Goethe University Frankfurt, Campus Riedberg, D-60438 Frankfurt am
Main, Germany}


\abstract{The mergers of binary compact objects such as neutron stars and black holes are of central interest to several areas of astrophysics, including as the progenitors of gamma-ray bursts (GRBs), sources of high-frequency gravitational waves and likely production sites for heavy element nucleosynthesis via rapid neutron capture (the $r$-process). These heavy elements include some of great geophysical, biological and cultural importance, such as thorium, iodine and gold. Here we present observations of the exceptionally bright gamma-ray burst GRB\,230307A. We show that GRB 230307A belongs to the class of long-duration gamma-ray bursts associated with compact object mergers, and contains a kilonova similar to AT2017gfo, associated with the gravitational-wave merger GW170817. We obtained James Webb Space Telescope mid-infrared (mid-IR) imaging and spectroscopy 29 and 61 days after the burst. The spectroscopy shows an emission line at 2.15 microns which we interpret as tellurium (atomic mass A=130), and a very red source, emitting most of its light in the mid-IR due to the production of lanthanides. These observations demonstrate that nucleosynthesis in GRBs can create $r$-process elements across a broad atomic mass range and play a central role in heavy element nucleosynthesis across the Universe.}




\maketitle

\begin{figure}
   \centering
   \includegraphics[width=13cm]{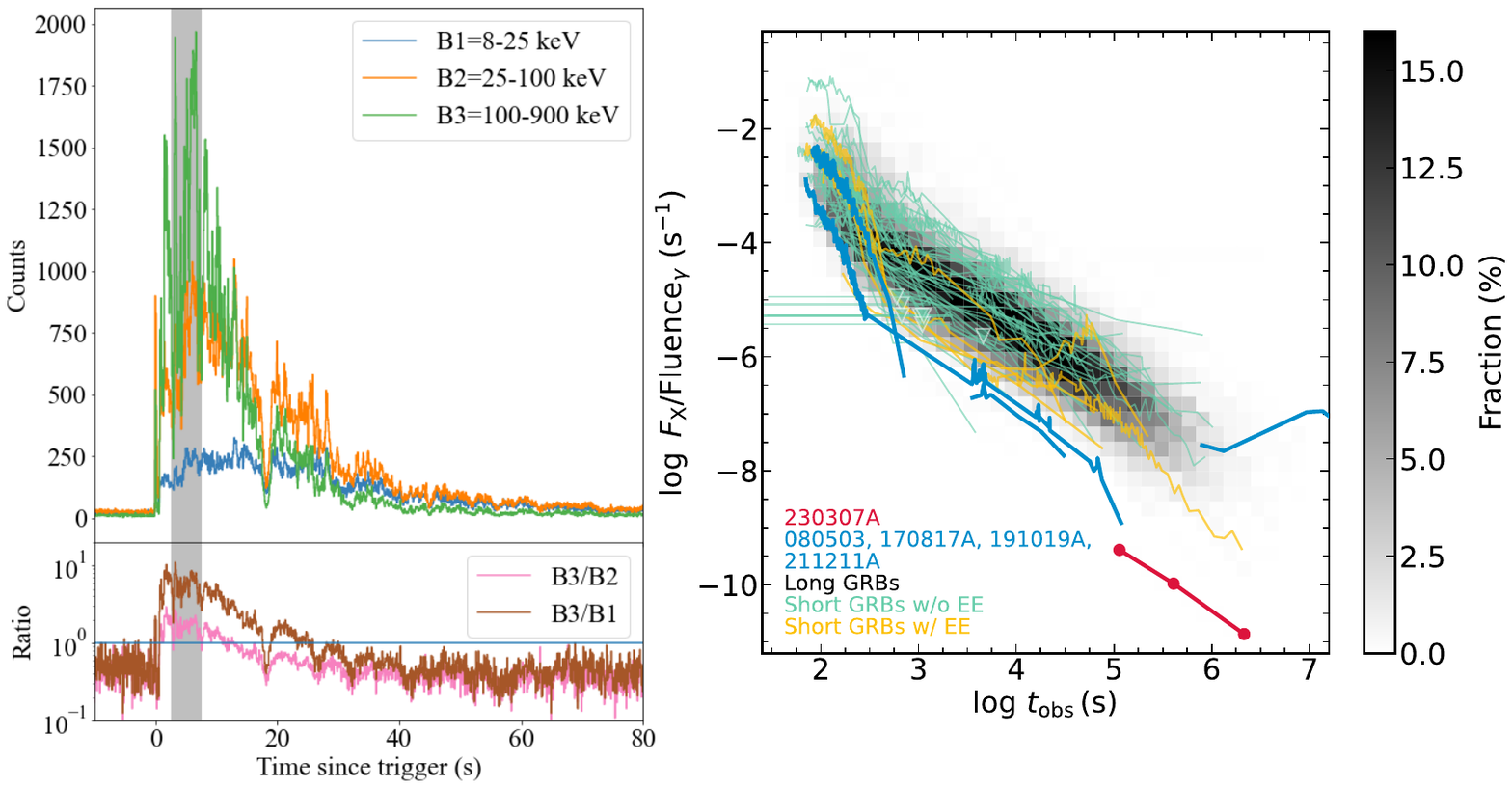}
      \caption{The high energy properties of GRB 230307A. The left panel shows the light curve of the GRB at 64 ms time resolution with the {\em Fermi}/GBM. The shaded region indicates the region where saturation may be an issue. The burst begins very hard, with the count rate dominated by photons in the hardest (100-900 keV) band, but rapidly softens, with the count rate in the hard band being progressively overtaken by softer bands (e.g. 8-25 keV and 25-100 keV) 
      beyond $\sim 20$s. This strong hard-to-soft evolution is reminiscent of GRB 211211A \citep{gompertz23} and is caused by the motion of two spectral breaks through the $\gamma$-ray regime (see Methods). The right panel shows the X-ray light curves of GRBs from the {\em Swift} X-ray telescope. These have been divided by the prompt fluence of the burst, which broadly scales with the X-ray light curve luminosity, resulting in a modest spread of afterglows. The greyscale background represents the ensemble of long GRBs. GRB 230307A is an extreme outlier of the $>1000$ {\em Swift}-GRBs, with an extremely faint afterglow for the brightness of its prompt emission. Other merger GRBs from long bursts occupy a similar region of parameter space. This suggests the prompt to afterglow fluence could be a valuable tool in distinguishing long GRBs from mergers and those from supernovae.  }\label{fig:GBM}
   \end{figure}

GRB 230307A was first detected
by the {\em Fermi} Gamma-ray Burst Monitor (GBM) at 15:44:06 UT on 7 Mar 2023 \citep{2023GCN.33405....1F}. It was an exceptionally bright gamma-ray burst 
with a duration of $T_{90} \sim 35$~s and a prompt fluence of (2.951 $\pm$  0.004) $\times 10^{-3}$\,erg\,cm$^{-2}$ in the 10-1000 keV band \citep{2023GCN.33411....1D} (Figure~\ref{fig:GBM}). These properties place this event at the peak of the distribution of the class of ``long-soft" GRBs. The measured fluence makes it the second-brightest GRB ever detected \citep{2023GCN.33414....B}.

In addition to {\em Fermi}, the burst was also detected by several other high-energy instruments (see Methods). The multiple detections enabled source triangulation by the InterPlanetary Network (IPN). The {\em Neil Gehrels Swift Observatory} (hereafter {\em Swift}) tiled the initial $\sim$2 sq. deg IPN region \citep{2022GCN.32641....1S} which revealed one candidate X-ray afterglow \citep{burrowsGCN} consistent with the final 8 sq. arcmin IPN localization \citep{2023GCN.33461....1K}. 

We obtained optical observations of the field of GRB 230307A with the ULTRACAM instrument, mounted on the 3.5m New Technology Telescope (NTT) at La Silla, Chile. Visual inspection of the images compared to those obtained with the Legacy Survey \citep{2019AJ....157..168D} revealed a new source coincident with the {\em Swift} X-ray source, and we identified it as the optical afterglow of GRB 230307A \citep{2023GCN.33439....1L}. This identification was subsequently confirmed via imaging by several additional observatories \cite{2023GCN.33441....1L,2023GCN.33447....1O,2023GCN.33449....1I}. The location was also serendipitously imaged by the Transiting Exoplanet Survey Satellite (TESS) from 3 days before to 3 days after the GRB \citep{fausnaugh23}.

\begin{figure*}
   \centering
   \includegraphics[width=\hsize]{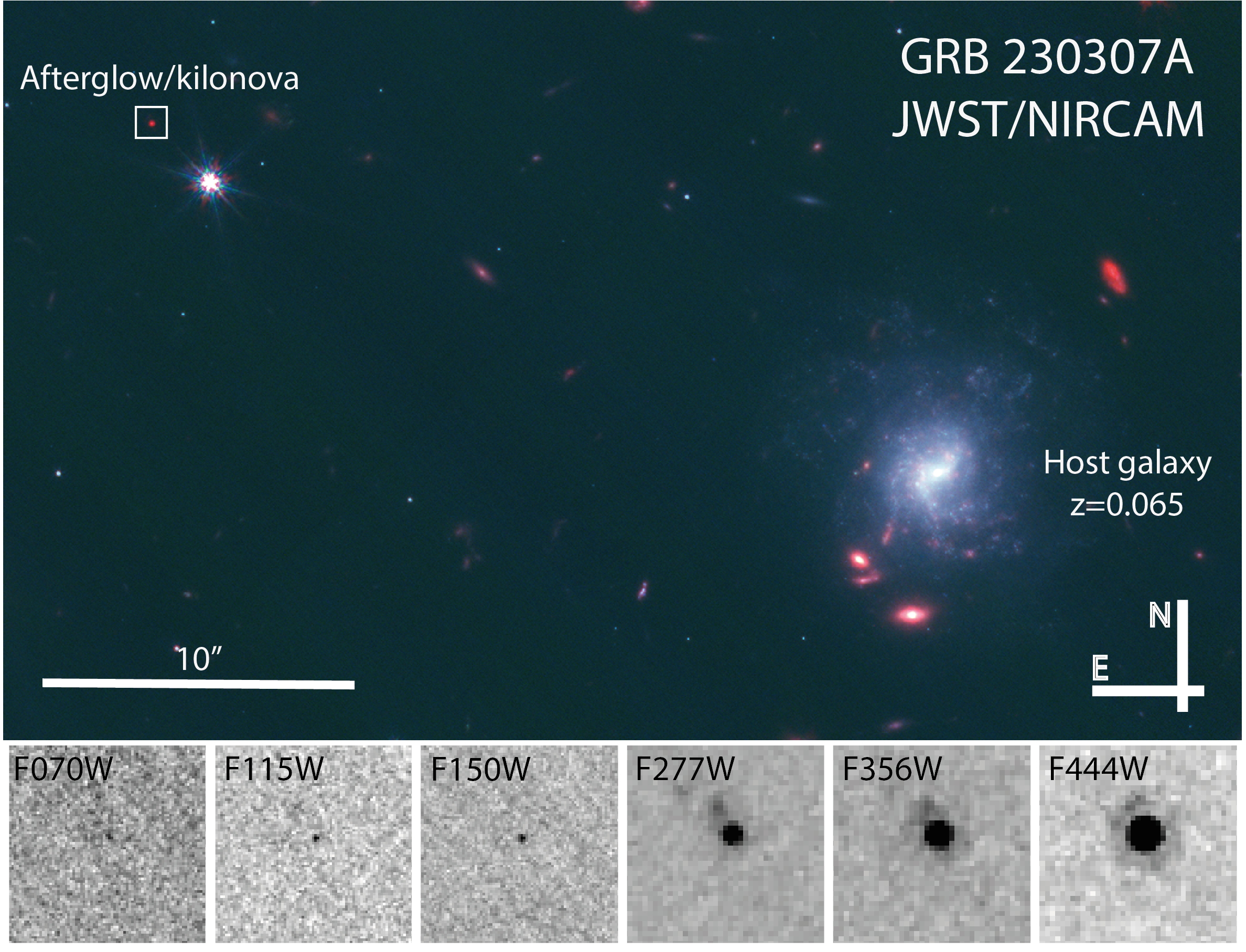}
      \caption{JWST images of GRB 230307A at 28.5 days post burst. The upper panel shows the wide field image combining the F115W, F150W and F444W images. The putative host  is the bright face-on spiral galaxy, while the afterglow appears at a 30-arcsecond offset, within the white box. The lower panel shows cut-outs of the NIRCAM data around the GRB afterglow location. The source is faint and barely detected in the bluer bands but very bright and well detected in the red. In the red bands, a faint galaxy is present northeast of the transient position. This galaxy has a redshift of $z=3.87$, but we consider it to be a background object unrelated to the GRB (see Supplementary Information).}\label{fig:jwst_finder}
   \end{figure*}

Following the precise localisation, we obtained extensive follow-up observations, in the optical and near infrared with the Gemini South Telescope and the Very Large Telescope (VLT); in the X-ray with the {\em Swift}/XRT and the Chandra X-ray observatory; and in the radio with the Australia Telescope Compact Array (ATCA) and MeerKAT in South Africa. This campaign included spectroscopy with the VLT X-shooter instrument, as well as the Multi Unit Spectroscopic Explorer (MUSE) integral field spectrograph. The latter provides redshift information for many galaxies in the field, including, in particular, a bright spiral galaxy at $z=0.0646 \pm 0.0001$ offset 30.2 arcseconds (38.9 kiloparsec in projection) from the burst position (Figure~\ref{fig:jwst_finder}). Of the optically detected galaxies in the field, this is the one with the lowest probability of being an unrelated chance alignment, and hence is most likely to be the host (see also \cite{gillandersGCN}). 

\begin{figure}
   \centering
   \includegraphics[width=\textwidth]{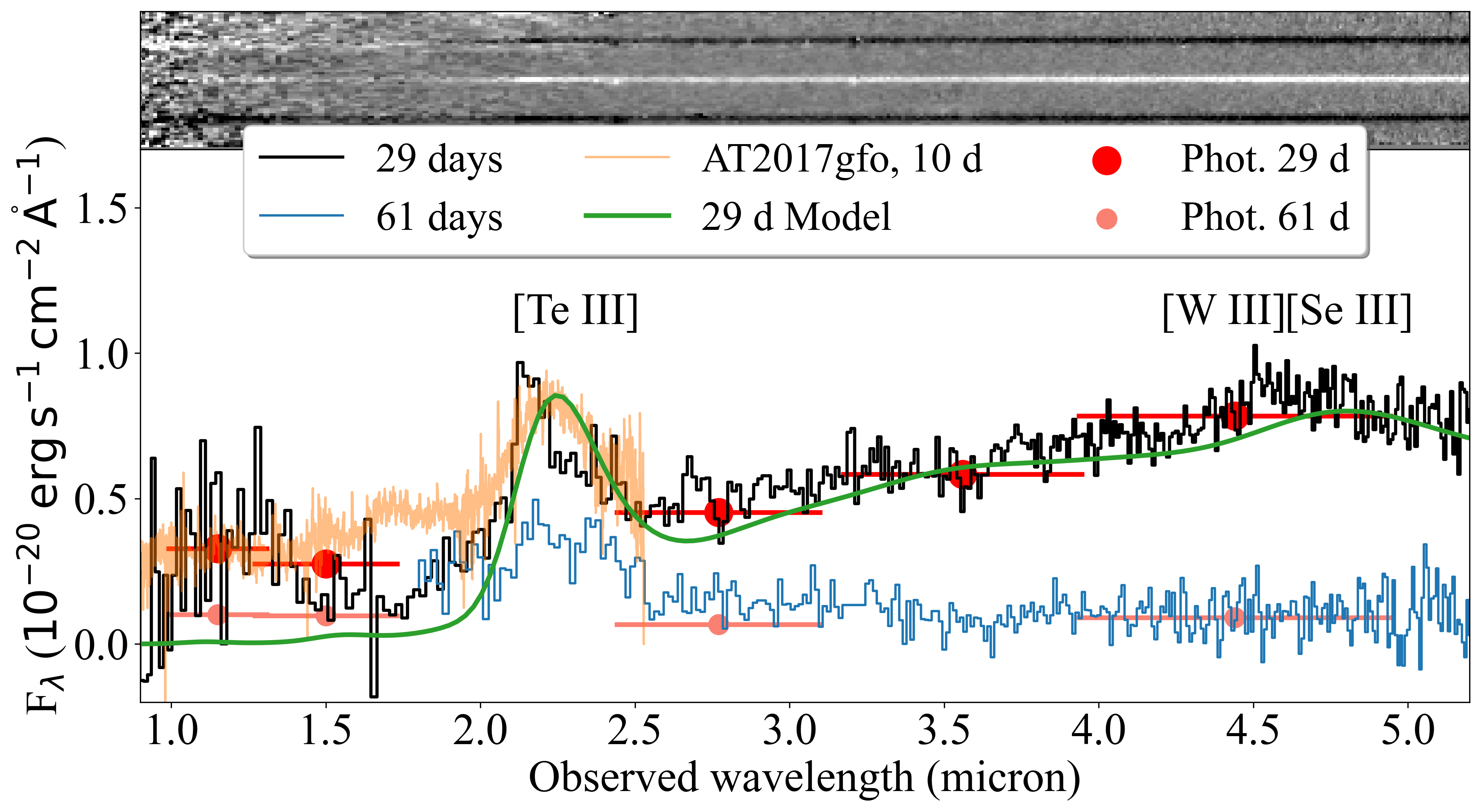}%
      \caption{JWST/NIRSpec spectroscopy of the counterpart of GRB 230307A taken on 5 April 2023. The top panel shows the 2-D spectrum rectified to a common wavelength scale. The transient is well detected beyond 2 microns but not short ward, indicative of an extremely red source. Emission lines from the nearby galaxy at $z=3.87$ can also be seen offset from the afterglow trace. The lower panel shows the 1D extraction of the spectrum in comparison with the latest (10-day) AT2017gfo epoch and different kilonova models. A clear emission feature can be seen at $\sim$ 2.15 microns at both 29 and 61 days. This feature is consistent with the expected location of [Te III], while redder features are compatible with lines from [Se III] and [W III]. This line is also clearly visible in the late time spectrum of AT2017gfo \cite{Gillanders23,hotokezaka23} }\label{NIRSpec}
   \end{figure}

Our ground-based campaign included imaging from 1.4 to 41 days after the burst (see Supplementary Information Table~\ref{tab:photometry}). At 11 days, infrared observations demonstrated a transition from an early blue spectral slope, to a much redder one with $i-K > 2.9$(AB). This extremely red colour appeared similar to the expectations for a kilonova, powered by the decay of unstable isotopes of heavy elements synthesised by rapid neutron capture within the ejecta produced during the merger of a neutron star and another compact object \cite{lattimer77,Li98,Metzger19}. Based on this detection, we requested James Webb Space Telescope (JWST) observations (GO 4434, 4445, PI Levan), which were initiated on 5 April 2023. At the first epoch (+28.9 days after GRB), we took 
6-colour observations with the Near Infrared Camera (NIRCam) (Figure~\ref{fig:jwst_finder}), as well as a spectrum with the Near Infrared Spectrograph (NIRSpec) covering $0.5-5.5$ microns (Figure~\ref{NIRSpec}).

The NIRCam observations reveal an extremely red source that is only weakly detected in the bluer bands, where F150W(AB) = $28.11 \pm 0.12$ mag, but rises sharply through the mid-IR to F444W(AB) = $24.4 \pm 0.01$ mag. The NIRSpec observations also exhibit this steep rise. A faint galaxy is detected in these data at $z=3.87$, offset approximately 0.3 arcseconds from the burst position. However, the burst's properties are inconsistent with an origin at this redshift, in particular because the implied isotropic equivalent energy release would exceed all known GRBs by an order of magnitude or more (see Supplementary Information). A second epoch of JWST observations was obtained approximately 61 days after the burst. These observations showed that the source had faded by 2.4 magnitudes in F444W, demonstrating a rapid decay expected in the kilonova scenario and effectively ruling out alternatives (see Supplementary Information).

\begin{figure}
   \centering
   \includegraphics[width=12cm]{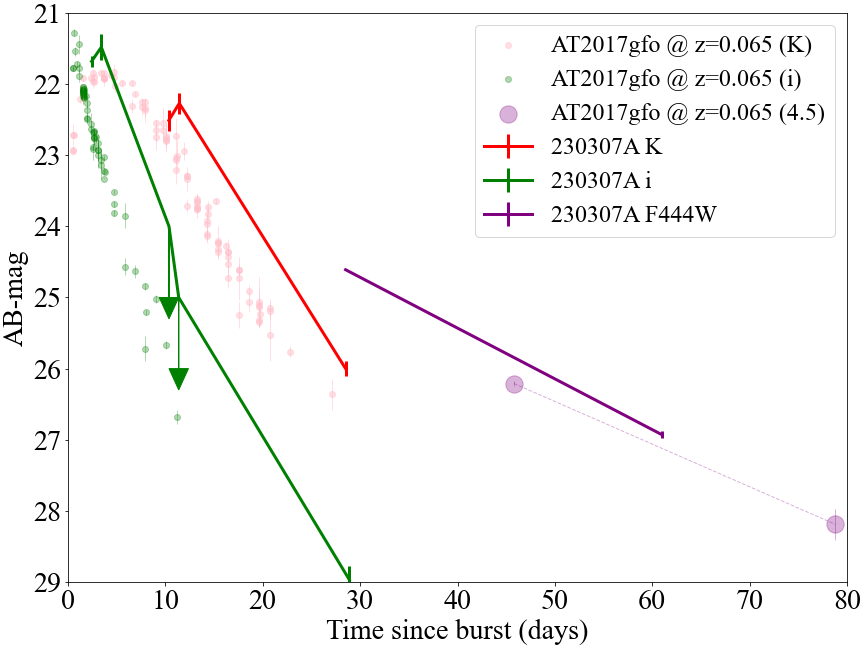}
      \caption{A comparison of the counterpart of GRB 230307A with AT2017gfo (the kilonova associated with GW170817) at $z=0.065$. Beyond $\sim 2$ days, the kilonova dominates the counterpart (see Supplementary Information). The decay rates in both the optical and IR are very similar to those in AT2017gfo. These are too rapid for any plausible afterglow model (e.g. as a power-law, they decay faster than $t^{-3.5}$ over a prolonged period). There is also good agreement in the late time slope between the measurements made at 4.4 microns with JWST and 4.5 microns for AT2017gfo with {\em Spitzer} \cite{kasliwal22}. }\label{fig:at2017gfo_comp}
   \end{figure}

Some of the burst properties are remarkably similar to those of the bright GRB 211211A, which was also shown to be accompanied by a kilonova \cite{rastinejad22,troja22,yang22}. In particular, the prompt emission consists of a hard pulse lasting for $\sim 19$ seconds, followed by much softer emission (Figure~\ref{fig:GBM}). The prompt emission spectrum is well modelled by a 
double broken power-law with two spectral breaks moving rapidly through the gamma-ray band (see Methods), suggesting a synchrotron origin of the emission \cite{gompertz23}. The X-ray afterglow is exceptionally faint, much fainter than most bursts when scaled by the prompt GRB fluence (see Figure~\ref{fig:GBM} and Supplementary Information). The development of the optical and IR counterpart is also similar to GRB 211211A, with an early blue colour and a subsequent transition to red on a timescale of a few days. In Figure~\ref{fig:at2017gfo_comp}, we plot the evolution of the counterpart compared with the kilonova AT2017gfo \cite{arcavi17,coulter17,pian17,smartt17,tanvir17,soares-santos17,valenti17,villar17}, identified in association with the gravitational-wave detected binary neutron star merger, GW170817 \cite{abbott_bns}. AT2017gfo is the most rapidly evolving thermal transient ever observed; much more rapid than supernovae or even, for example, fast blue optical transients \cite{drout14}. The counterpart of GRB 230307A appears to show near identical decline rates to AT2017gfo both at early times in the optical and IR, and later in the mid-IR \cite{kasliwal22}. These similarities are confirmed by a joint fit of afterglow and kilonova models to our multi-wavelength data (see Supplementary Information). Our JWST observations rule out supernovae: for any redshift $z<1$, a supernova would need to be more than $100$ times fainter than the canonical GRB-supernova, SN 1998bw, to be compatible with our observations. Therefore, we conclude that GRB 230307A is a long-duration GRB formed from a compact object merger. This falls into a class that includes GRB 211211A \cite{rastinejad22,troja22,yang22,mei22}, GRB 060614 \cite{gehrels06}, GRB 111005A \cite{michalowski18} and GRB 191019A \cite{levan23b}, among others.

The JWST observations provide a view of a kilonova in the mid-IR with high spatial resolution and sensitivity. On timescales of $\sim 30$ days, it is apparent that the kilonova emits almost all of its light in the mid-infrared, beyond the limits of sensitive ground-based observations (effectively limited to below 2.5 microns). This is consistent with previous model predictions \cite{wollaeger18}, but has not previously been observationally confirmed. Late-time studies of such emission in the nebular phase must therefore be conducted in the mid-IR. Strikingly, despite its powerful and long-lived prompt emission that stands in stark contrast to GRB 170817A, the GRB 230307A kilonova is remarkably similar to AT~2017gfo. This was also the case for the long-lived merger GRB 211211A \cite{rastinejad22,troja22,yang22,mei22}, and suggests the kilonova signal, particularly the red component, is relatively insensitive to the GRB.


Our NIRSpec spectrum shows a broad emission feature with a central wavelength of 2.15 microns, visible in both epochs of JWST spectroscopy. At longer wavelengths, the spectrum displays a slowly rising continuum up to 4.5 microns followed by either an additional feature or change of spectral slope. The colours of the counterpart at this time can be readily explained by kilonova models (see Supplementary Information Section~\ref{sec:modeling}).

A similar emission-like feature is also visible in the later epochs of X-shooter observations of AT2017gfo \cite{pian17}, measured at 2.1\,microns by \cite{Gillanders23}.
This further strengthens both the kilonova interpretation and the redshift measurement of GRB 230307A (Figure \ref{NIRSpec}). We interpret this feature as arising  from the forbidden [Te III] transition between the ground level and the first fine structure level of tellurium, with an experimentally-determined wavelength of 2.1044 microns \cite{Joshi1992}. 
The presence of tellurium is plausible, as it lies at the second peak in the $r$-process abundance pattern, which occurs at atomic masses around $A \approx 130$ \cite{hotokezaka18}. It should therefore be abundantly produced in kilonovae, as is seen in hydrodynamical simulations of binary neutron star mergers with nucleosynthetic compositions similar to those favoured for AT2017gfo \cite{Gillanders22}. Furthermore, the typical ionization state of Te in kilonova ejecta is expected to be Te III at this epoch because of the efficient radioactive ionization \cite{Pognan2022MNRAS}.  Tellurium has recently been suggested as the origin of the same feature in the spectrum of AT2017gfo \cite{hotokezaka23}. A previous study \cite{Gillanders23} also identified this tellurium transition and noted that the observed feature is most likely two blended emission lines. However, alternative transitions from heavy $r$-process elements have been considered for this feature \citep[e.g. Ce III;][]{Gillanders23}. Tellurium can also be produced via the slower capture of neutrons in the $s$-process. Indeed, this line is also seen in planetary nebulae \cite{Madonna2018}. The detection of [Te III] $2.1\,{\rm \mu m}$ extends on the earlier detection of strontium, a first $r$-process peak element, in the early time photospheric emission of AT2017gfo \cite{watson19}. 
The mass of Te III estimated from the observed line flux is $\sim 10^{-3}M_{\odot}$ (see Supplementary Information \ref{sec:spec_model}). Although weaker, we also note that the spectral feature visible at 4.5 microns is approximately consistent with the expected location of the first peak element selenium and the near-third peak element tungsten \cite{hotokezaka22}. 

Detailed spectral fitting at late epochs is challenging because of the breakdown of the assumptions regarding local thermodynamic equilibrium (LTE) which are used to predict kilonova spectra at earlier ages, as well as fundamental uncertainties in the atomic physics and electron transitions in the highly complex electron shells of $r$-process elements. However, these observations provide a calibration sample for informing future models. The red continuum emission indicates a large opacity in the mid-IR at low temperatures, e.g., $\sim 10\,{\rm cm^2g^{-1}}$ at $\sim 700\,{\rm K}$. Since the Planck mean opacity under LTE is expected to be less than $1\,{\rm cm^2g^{-1}}$ at $\lesssim 1000\,{\rm K}$ \cite{tanaka20}, this large opacity may suggest that lanthanides are abundant in the ejecta. 
Indeed, it has been shown that non-LTE effects can increase the lanthanide opacity in mid-IR at late times \cite{pognan22b}. 
Therefore, systematic studies of non-LTE opacity are necessary to answer the question whether lanthanides are the origin of the red emission at this epoch. 

A fit to our combined MUSE + JWST data for the host galaxy (Supplementary Information) suggests a relatively low mass ($\sim 2.5 \times 10^9$ M$_{\odot}$)  dominated by an older $\sim 10$ Gyr population, but with a second more recent burst of star formation. These properties are entirely consistent with the population of short GRB host galaxies \cite{oconnor22,nugent22}. The host normalised offset of the burst from the host galaxy places it in the top 10\% of those seen for short GRBs \cite{fong22,oconnor22}. The offset could readily be achieved by a binary with a velocity of a few hundred km s$^{-1}$ and a merger time of hundreds of millions of years.
Alternatively, in the second epoch of JWST observations, a faint source is  detected in the F150W images at the location of the transient. This may represent continued emission from the transient. However, its absolute magnitude of M$_{F150W}$ $\sim -8.5$ is comparable to the absolute magnitude of globular clusters in which dynamical interactions could be at play to create merging systems at enhanced rates \cite{grindlay06}. Future observations should readily distinguish between a fading afterglow or underlying cluster.

It is striking that GRB 230307A is an extremely bright GRB, with only the exceptional GRB 221009A being brighter \citep{burns23}. 
Of the ten most fluent {\em Fermi}/GBM GRBs, two are now associated with kilonovae (230307A and 211211A), three are associated with supernovae, and the nature of the remaining five appears unclear (see Table~\ref{tab:bright_bursts}). For bright GRBs, there may be a significant contribution from mergers. Indeed, such a conclusion can also be reached by considering the energetics. Both GRB 230307A and GRB 211211A have isotropic equivalent energies of $E_\mathrm{iso} > 10^{51}$ erg. The majority of local GRBs for which the connection between GRBs and core-collapse supernovae has been established are much less energetic (typically $E_\mathrm{iso} > 10^{49-50}$ erg) and it has been suggested that they represent a separate population powered via different processes \cite{virgili09}. For more energetic bursts in the local Universe (where supernovae can still readily be discovered) the fraction of long GRBs with and without supernovae appear similar (see Supplementary Information). If a substantial number of long GRBs are associated with compact object mergers, they provide an essential complement to gravitational-wave (GW) detections. Firstly, joint GW-GRB detections, including long GRBs, can push the effective horizons of GW detectors to greater distances and provide much smaller localisations. In the case of GRB 230307A, the distance of 300 Mpc could have provided a robust detection in gravitational waves for the relevant O4 sensitivity \cite{rastinejad22,sarin23}. Secondly, long GRBs can be detected without GW detectors, providing a valuable route to enhancing the number of kilonova detections. Thirdly, JWST can detect kilonova emission at redshifts substantially beyond the horizons of the current generation of GW detectors, enabling the study of kilonovae across a greater volume of the Universe.

The duration of the prompt $\gamma$-ray emission in these mergers remains a challenge to explain. In particular, the natural timescales for emission in compact object mergers are much shorter than the measured duration of GRB 230307A. Previously suggested models that may also explain GRB 230307A include magnetars \cite{metzger08,bucciantini12}, black hole - neutron star mergers \cite{rosswog07,desai19}, or even neutron star - white dwarf systems \cite{yang22,zhong23}. Recent results have also shown that the jet timescale does not directly track the accretion timescale in compact object mergers, and that long GRBs may be created from very short lived engines \cite{gottlieb23}, and hence from binary neutron star mergers without magnetars.

There is evidence that the kilonova in GRB 230307A produced elements across a wide range of atomic mass. The detection of second peak elements in the spectrum of a kilonova demonstrates that nuclei with atomic masses around $A \sim 130$ are being created in the mergers of compact objects. Many second peak elements have important biological roles. For example iodine is essential for mammals and may have been used by the single cell Last Universal Common Ancestor \cite{crockford09}. The creation of these elements in compact object mergers, which can have long delay times, may have important consequences for the time at which certain evolutionary channels become plausible.

\clearpage

\clearpage

\section{Methods}\label{sec2}

\section{Observations}
Below we outline the observational data that were used in this paper. Magnitudes are given in the AB system unless stated otherwise. We utilize cosmology resulting from the Planck observations \cite{planck18}. All uncertainties are given at the 1$\sigma$ level unless explicitly stated. 

\subsection{\texorpdfstring{$\gamma$-ray}{gamma-ray} observations}
GRB 230307A was first detected by {\em Fermi}/GBM and GECAM at 15:44:06 UT on 7 Mar 2023 \citep{2023GCN.33405....1F}. It had a duration of $T_{90} \sim 35$s and an exceptionally bright prompt fluence of (2.951 $\pm$  0.004) $\times 10^{-3}$\,erg\,cm$^{-2}$ \citep{2023GCN.33411....1D}. The burst fell outside of the coded field-of-view of the {\em Swift} Burst Alert Telescope (BAT), and so did not receive a sub-degree localisation despite a strong detection. However, detections by {\em Swift}, GECAM \citep{2023GCN.33406....1X}, STIX on the Solar Orbiter 
\citep{gcn33410}, AGILE \citep{2023GCN.33412....1C}, ASTROSAT \citep{2023GCN.33415....1N}, GRBalpha \citep{2023GCN.33418....1D}, VZLUSAT \citep{2023GCN.33424....1R}, Konus-WIND \citep{2023GCN.33427....1S} and ASO-HXI \citep{2023GCN.33438....1L} enabled an enhanced position via the InterPlanetary Network to increasingly precise localisations of 1.948 deg$^2$ \citep{2023GCN.33413....1K}, 30 arcmin$^2$ \citep{2023GCN.33425....1K}, and ultimately to 8 arcmin$^2$ \citep{2023GCN.33461....1K}. This was sufficiently small to enable tiling with {\em Swift} and ground-based telescopes.



\subsubsection{{\em Fermi}/GBM data analysis}\label{gbm_analysis}
In Figure~\ref{fig:GBM}, we plot the light curve of GRB 230307A as seen by the {\em Fermi}/GBM in several bands, built by selecting Time Tagged Event (TTE) data, binned with a time resolution of 64 ms. The highlighted time interval of 3--7~s after trigger are affected by data loss due to the bandwidth limit for TTE data \cite{gcn33551}.

For the spectral analysis, we made use of the CSPEC data, which have 1024 ms time resolution. Data files were obtained from the online archive\footnote{\url{https://heasarc.gsfc.nasa.gov/W3Browse/fermi/fermigbrst.html}}. Following the suggestion reported by the {\em Fermi} Collaboration \citep{gcn33551}, we analysed the data detected by NaI 10 and BGO 1, which had a source viewing angle less than 60$^{\circ}$, and excluded the time intervals affected by pulse pile-up issues (from 2.5 s to 7.5 s). The data extraction was performed with the public software {\sc gtburst}, while data were analysed with {\sc Xspec}. The background, whose time intervals have been selected before and after the source, was modelled with a polynomial function whose order is automatically found by {\sc gtburst} and manually checked. In the fitting procedure, we used inter--calibration factors among the detectors, scaled to the only NaI analysed and free to vary within 30\%. We used the PG-Statistic, valid for Poisson data with a Gaussian background. The best-fit parameters and their uncertainties were estimated through a Markov Chain Monte Carlo (MCMC) approach. We selected the time intervals before and after the excluded period of 2.5-7.5 s due to instrumental effects. In particular, we extracted 2 time intervals from 0 to 2.5 s (1.25 s each) and 14 time intervals from 7.5 s to 40.5 (bin width of 2 s, except the last two with integration of 5 s to increase the signal-to-noise ratio), for a total of 16 time intervals. We fitted the corresponding spectra with the two smoothly broken power-law (2SBPL) function \cite{ravasio18,ravasio19}, which has been shown to successfully model the synchrotron-like spectral shape of bright long GRBs, including the merger-driven GRB 211211A \cite{gompertz23}. 

From our spectral analysis we found that all spectra up to $\sim$ 20 s are well modelled by the 2SBPL function, namely they are described by the presence of two spectral breaks inside the GBM band (8 keV--40 MeV). 
In particular, in the time intervals between 7.5 and 19.5 s, the low-energy break $E_\mathrm{break}$ is coherently decreasing from $304.3_{-2.6}^{+5.2}$ keV down to $52.1_{-5.1}^{+4.3}$ keV, and the typical $\nu F_{\nu}$ peak energy $E_\mathrm{peak}$ is also becoming softer, moving from $\sim$ 1 MeV to 450 keV. The spectral indices of the two power-laws below and above the low-energy break are distributed around the values of -0.82 and -1.72, which are similar to the predictions for synchrotron emission in marginally fast-cooling regime (i.e. $-2/3$ and $-3/2$). This is consistent with what has been found in GRB 211211A \cite{gompertz23}. 
We notice, however, that in all spectra the high-energy power-law above $E_\mathrm{peak}$ is characterised by a much softer index (with a mean value of $-4.10 \pm 0.24$) with respect to the  value of $\sim$ -2.5  typically found in {\em Fermi} GRBs. This suggests that the spectral data might require a cut-off at high energy, although further investigations are needed to support this.
From 19.5 s until 40.5 s (the last time interval analysed), all the break energies are found to be below 20 keV, close to the GBM low energy threshold.  In the same time intervals, the peak energy $E_\mathrm{peak}$ decreases from $682.4_{-6.1}^{+3.2}$ to $123.1_{-4.9}^{+5.4}$ keV, and the index of the power-law below the peak energy is fully consistent (mean value of $-1.45 \pm 0.06$) with the synchrotron predicted value of -1.5.

\subsection{Optical observations}

\subsubsection{NTT - Afterglow discovery}

Following the refinement of the IPN error box to an area of 30 arcmin$^2$ \citep{2023GCN.33425....1K}, we obtained observations of the field of GRB 230307A with the ULTRACAM instrument \citep{dhillon07}, mounted on the 3.5m New Technology Telescope (NTT) at La Silla, Chile. The instrument obtains images in 3 simultaneous bands, and is optimised for short exposure, low dead-time observations \citep{dhillon07}. We obtained $10 \times 20$ s exposures in two pointings in each of the Super SDSS $u$, $g$ and $r$, bands (where the Super SDSS bands match the wavelength range of the traditional SDSS filters but with a higher throughput; \citealt{dhillon21}). The observations began at 01:53:21 UT on 2023-03-09, approximately 34 hr after the GRB. The images were reduced via the HIPERCAM pipeline \citep{dhillon21} using bias and flat frames taken on the same night. Visual inspection of the images compared to those obtained with the Legacy Survey \citep{2019AJ....157..168D} revealed a new source coincident with an X-ray source identified via {\em Swift}/XRT observations \cite{burrowsGCN}, and we identified it as the likely optical afterglow of GRB 230307A \citep{2023GCN.33439....1L}. The best available optical position of this source (ultimately measured from our {\em JWST} observations, see below) is RA(J2000) = 04:03:26.02, Dec(J2000) = $-$75:22:42.76, with an uncertainty of 0.05 arcseconds in each axis. The IPN error box and the footprint of the ULTRACAM observations are shown in Figure~\ref{fig:ultracam_finder}.

This identification was subsequently confirmed via observations from a number of additional observatories, including \cite{2023GCN.33441....1L,2023GCN.33447....1O,2023GCN.33449....1I,2023GCN.33459....1B,2023GCN.33471....1B,2023GCN.33485....1G}. We acquired two further epochs of observations with ULTRACAM on the following nights with $10 \times 20$s exposures in the Super SDSS $u$, $g$ and $i$, bands. Aperture photometry of the source is reported in Table~\ref{tab:photometry}, and is reported relative to the Legacy survey for the $gri$ bands, and to SkyMapper for the $u$-band. 

\subsubsection{TESS}

The prompt and afterglow emission of GRB 230307A was detected by TESS, which observed the field continuously from 3 days before the {\em Fermi} trigger to 3 days after at a cadence of 200\,s \citep{fausnaugh23}. A reference image was subtracted from the observations to obtain GRB-only flux over this period. The measured flux in the broad TESS filter (600nm -- 1000nm) is corrected for Galactic extinction and converted to the $I_{\rm c}$
band assuming a power-law spectrum with $F \propto \nu^{-0.8}$. We then bin the light curve logarithmically, taking the mean flux of the observations in each bin and converting to AB magnitudes. A systematic error of 0.1 magnitudes was added in quadrature to the measured statistical errors to account for the uncertainties in the data processing. These data are presented in Table~\ref{tab:photometry}.

\subsubsection{{\em Swift}/UVOT}
The {\em Swift} Ultra-violet/Optical Telescope \citep[UVOT; ][]{roming} began observing the field of GRB 230307A
$\sim84.6$\,ks after the {\em Fermi}/GBM trigger \citep{2023GCN.33405....1F}. The source counts were extracted using a source region of 5 arcsec radius.
Background counts were extracted using a circular region of 20 arcsec radius located in a source-free part of the sky.
The count rates were obtained from the image lists using the {\em Swift} tool \textsc{uvotsource}.
A faint catalogued unrelated source also falls within the 5 arcsec radius, this will affect the photometry, particularly at late times.
We, therefore, requested a deep template image in $white$ in order to estimate the level of contamination. We
extracted the count rate in the template image using the same 5 arcsec radii aperture. This was subtracted from the source count rates to obtain the afterglow count rates. The afterglow count rates were converted to magnitudes using the UVOT photometric zero points \citep{poole,bre11}.

\subsubsection{Gemini}
We obtained three epochs of K-band observations using the FLAMINGOS-2 instrument on the Gemini-South telescope. These observations were reduced through the DRAGONS pipeline to produce dark and sky-subtracted and flat-fielded images \citep{DRAGONS19}. At the location of the optical counterpart to GRB 230307A, we identify a relatively bright K-band source in the first and second epochs, with only a upper limit in epoch 3. We report our photometry, performed relative to secondary standards in the VISTA hemisphere survey \citep{vhs}, in Table~\ref{tab:photometry}. 

\subsubsection{VLT imaging}
\begin{figure*}
   \centering
   \includegraphics[width=\hsize]{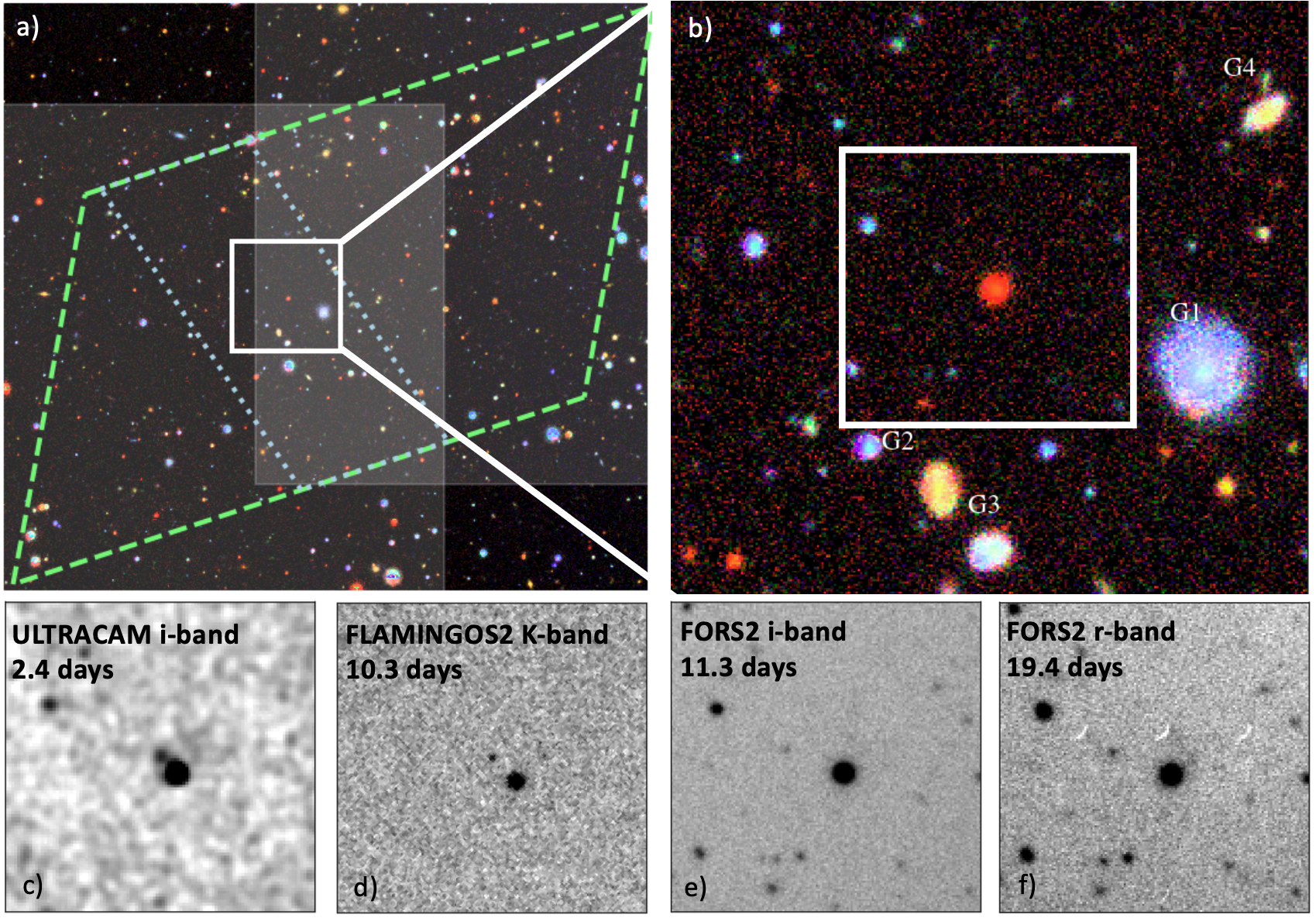}
      \caption{Ground based optical and IR imaging of the afterglow of GRB 230307A. Panel (a) shows a legacy survey image of the field, the green trapezoid shows the IPN error box at the time of afterlgow discovery, with the blue dotted lines showing its ultimate refinement. The shaded regions represent observations taken with ULTRACAM 1.4 days post burst. Panel (b) shows the same image, zoomed in to a region around the afterglow. The red source in the centre is an unrelated foreground star (confirmed both by its measured proper motion and spectrum), and several other galaxies can also be seen. Panel (c) shows an ULTRACAM image, where the afterglow can be seen to the north-east of the star. The remaining panels (d,e,f) show the afterglow imaged on $\sim 10-20$ day timescales. At 10 days the source is undetected in deep FORS2 i-band imaging, but well detected in the K-band with Gemini-South. This very red colour was suggestive of an additional component over and above any afterglow, and motivated further follow-up.}\label{fig:ultracam_finder}
   \end{figure*}

We carried out observations of the GRB 230307A field with the 8.2-m VLT telescopes located in Cerro Paranal, Chile. The observations were obtained with the FORS2 camera (mounted on the Unit Telescope 1, UT1, ANTU) in $B$, $R$, $I$ and $z$ bands at multiple epochs, and with the HAWK-I instrument (mounted on the Unit Telescope 4, UT4, Yepun) in the $K$ band at one epoch. All images were reduced using the standard ESO (European Southern Observatory) Reflex pipeline \citep{2013A&A...559A..96F}. The source was detected in the FORS2 $z$-band image at $\sim$6.4 days after the {\em Fermi}/GBM detection. A single $r'$-band observation of the GRB 230307A was also executed with the 2.6m VLT Survey Telescope (VST) after 2.37 days from the GRB discovery. In later observations the source was not detected (see bottom right panels of Fig.~\ref{fig:ultracam_finder}) and the upper limit values at 3$\sigma$ level are reported in Table~\ref{tab:photometry}.

\subsubsection{VLT spectroscopy}
To attempt to measure the redshift of GRB 230307A and the nearby candidate host galaxies, we obtained spectroscopy with the VLT utilising both the X-shooter and MUSE instruments, mounted respectively on the Unit Telescope 2 (UT2, Kueyen) and on UT4 (Yepun). 

X-shooter spectroscopy, covering the wavelength range 3000--22000 \AA\ was undertaken on 2023-03-15. Observations were taken at a fixed position angle, with the slit centred on a nearby bright star. 
X-shooter data have been reduced with standard {\em esorex} recipes. Given that only two of the four nod exposures were covering the GRB position, resulting in a total exposure time of 2400s on-source, we have reduced each single exposure using the stare mode data reduction. Then, we have stacked the two 2D frames covering the GRB position using dedicated post-processing tools developed in a python framework \cite{Selsing19}.
We further obtained observations with the MUSE integral field unit on 2023-03-23. The MUSE observations cover multiple galaxies within the field, as well as the GRB position, and cover the wavelength range 4750-9350 \AA. MUSE data have been reduced using standard {\em esorex} recipes embedded within a single python script that performs the entire data reduction procedure. Later, the resulting datacube has been corrected for sky emission residuals using ZAP \cite{Soto2016}. The MUSE observations reveal the redshifts for a large number of galaxies in the field, including a prominent spiral G1 at $z=0.0646$ \citep[see also][]{2023GCN.33485....1G} and a group of galaxies, G2, G3 and G4 at $z=0.263$.


\subsection{X-ray afterglow}

{\em Swift} began tiled observations of the IPN localisation region with its X-ray Telescope \citep[XRT;][]{burrows05} at 12:56:42 on 8 Mar 2023\footnote{\href{https://www.swift.ac.uk/xrt_products/TILED_GRB00110/}{https://www.swift.ac.uk/xrt\_products/TILED\_GRB00110/}} \citep{2023GCN.33419....1E}. XRT made the first reported detection of the afterglow (initially identified as `Source 2') with a count rate of $0.019 \pm 0.004$\,cts\,$^{-1}$ \citep{2023GCN.33429....1B} and later confirmed it to be fading with a temporal power-law index of $1.1^{+0.6}_{-0.5}$ \citep{2023GCN.33465....B}. XRT data were downloaded from the UK {\em Swift} Science Data Centre \citep[UKSSDC;][]{evans07,evans09}. 

We further obtained observations with the {\em Chandra} X-ray observatory (programme ID 402458: PI Fong/Gompertz). A total of 50.26 ks (49.67 ks of effective exposure) of data were obtained in three visits between 31 March 2023 and 2 April 2023. The source was placed at the default aim point on the S3 chip of the ACIS detector. At the location of the optical and X-ray afterglow of GRB 230307A, we detect a total of 12 counts, with an expected background of $\sim 1$, corresponding to a detection of the afterglow at $>5 \sigma$ based on the photon statistics of \citep{kraft91}. To obtain fluxes, we performed a joint spectral fit of the {\em Chandra} and {\em Swift}/XRT data. The best fitting spectrum, adopting uniform priors on all parameters, is a power law with a photon index of $\Gamma=2.50^{+0.30}_{-0.29}$ when fitting with a Galactic $N_H = 1.26 \times 10^{21}$ cm$^{-2}$ \cite{willingale13} and zero intrinsic absorption (neither XRT nor {\em Chandra} spectra have sufficient signal to noise to constrain any intrinsic absorption component). The resultant flux in the 0.3 -- 10 keV band is $F_\mathrm{X}(1.7\,\mathrm{d}) = 4.91_{-0.79}^{+0.89}\times 10^{-13}$ erg cm$^{-2}$ s$^{-1}$ during the XRT observation and $F_\mathrm{X}(24.8\,\mathrm{d}) = 1.19_{-0.62}^{+0.87}\times 10^{-14}$ erg cm$^{-2}$ s$^{-1}$ during the Chandra observation. Due to the low count number, the \emph{Chandra} flux posterior support extends to considerably below the reported median, with the 5th percentile being as low as $F_\mathrm{X,5th} = 3\times 10^{-15}$ erg cm$^{-2}$ s$^{-1}$. If a uniform-in-the-logarithm prior on the flux were adopted, this would extend to even lower values. \emph{Chandra} and XRT fluxes are converted to 1\,keV flux densities using the best fit spectrum (Table~\ref{tab:radio_xrays}). 

\subsection{ATCA}

Following the identification of the optical afterglow \citep{levan2023}, we requested target of opportunity observations of GRB 230307A (proposal identification CX529) with the Australia Telescope Compact Array (ATCA) to search for a radio counterpart. 
These data were processed using {\sc Miriad} \cite{Sault1995}, which is the native reduction software package for ATCA data using standard techniques. Flux and bandpass calibration were performed using PKS\,1934--638, with phase calibration using interleaved observations of 0454-810.

The first observation took place on 2023-03-12 at 4.46\,d post-burst, which was conducted using the 4\,cm dual receiver with frequencies centered at 5.5 and 9\,GHz, each with a 2\,GHz bandwidth. The array was in the 750C configuration\footnote{https://www.narrabri.atnf.csiro.au/operations/array\_configurations/configurations.html} with a maximum baseline of 6\,km. A radio source was detected at the position of the optical afterglow at 9\,GHz with a flux density of $92 \pm 22\mu$Jy but went undetected at 5.5\,GHz ($3\sigma$ upper limit of $84\mu$Jy). 
A further two follow-up observations were also obtained swapping between the 4\,cm and 15\,mm dual receivers (the latter with central frequencies of 16.7 and 21.2\,GHz, each with a 2\,GHz bandwidth).
During our second epoch at 10.66\,d we detected the radio counterpart again, having become detectable at 5.5\,GHz with marginal fading at 9\,GHz. By the third epoch, the radio afterglow had faded below detectability. We did not detect the radio transient at 16.7 or 21.2\,GHz in either epoch.
All ATCA flux densities are listed in Table~\ref{tab:radio_xrays}.

\subsection{MeerKAT}

We were awarded time to observe the position of GRB 230307A with the MeerKAT radio telescope via a successful Director’s Discretionary Time proposal (PI: Rhodes, DDT-20230313-LR-01). The MeerKAT radio telescope is a 64-dish interferometer based in the Karoo Desert, Northern Cape, South Africa \citep{Jonas2016}. Each dish is 12\,m in diameter and the longest baseline is $\sim$8\,km allowing for an angular resolution of $\sim7$\,arccsec and a field of view of 1\,deg$^{2}$. The observations we were awarded were made at both L and S-band.

GRB 230307A was observed over three separate epochs between seven and 41\,days post-burst. The first two observations were made at both L and S4-band (the highest frequency of the five S-band sub-bands), centred at 1.28 and 3.06\,GHz with a bandwidth of 0.856 and 0.875\,GHz, respectively. Each observation spent two hours at L-band and 20 minutes at S4-band. The final observation was made only at S4-band with one hour on target. Please see the paper by MPIfR for further details on the new MeerKAT S-band receiver. 

Each observation was processed using \textsc{oxkat}, a series of semi-automated Python scripts designed specifically to deal with MeerKAT imaging data \citep{heywood2020}. The scripts average the data and perform flagging on the calibrators from which delay, bandpass and gain corrections are calculated and then applied to the target. The sources J0408-6545 and J0252-7104 were used at the flux and complex gain calibrator, respectively. Flagging and imaging of the target field are performed. We also perform a single round of phase-only self-calibration. We do not detect a radio counterpart in any epoch in either band. The rms noise in the field was measured using an empty region of the sky and used to calculate 3$\sigma$ upper limits which are given in Table \ref{tab:radio_xrays}.

\subsection{JWST observations}
We obtained two epochs of observations of the location of GRB 230307A with JWST. The first on 5 April 2023, with observations beginning at 
00:16 UT (MJD=60039.01), 28.4 days after the burst (under programme GO 4434, PI Levan), and the second on 8 May 2023, 61.5 days after the burst (programme 4445, PI Levan). The observations were at a post-peak epoch because the source was not in the JWST field of regard at the time of the burst and only entered it on 2 April 2023. 

At the first epoch, we obtained observations in the F070W, F115W, F150W, F277W, F356W and F444W filters of NIRCam \citep{beichman12}, as well as a prism spectrum with NIRSpec \citep{jakobsen22}. In the second epoch we obtained NIRCam observations in F115W, F150W, F277W and F444W and a further NIRSpec prism observation. However, in the second epoch the prism observation is contaminated by light from the diffraction spike of a nearby star and is of limited use, in particular at the blue end of the spectrum. We therefore use only light redwad of 1.8 microns. However, even here we should be cautious in interpreting the overall spectral shape. However, the feature at 2.15 microns is visible in both the 29 and 61 day spectra. 

We reprocessed and re-drizzled the NIRCam data products to remove $1/f$ striping and aid point spread function recovery, with the final images having plate scales of 0.02 arcsec/pixel (blue channel) and 0.04 arcsec/pixel (red channel). 

In the NIRCam imaging we detect a source at the location of the optical counterpart of GRB 230307A. This source is weakly detected in all three bluer filters (F070W, F115W and F150W), but is at high signal-to-noise ratio in the redder channels (see Figure~\ref{fig:jwst_finder}). The source is compact. We also identify a second source offset (H1) approximately 0.3 arcseconds from the burst location. This source is also weakly, or non-detected in the bluer bands, and is brightest in the F277W filter.  

Because of the proximity of the nearby star and a contribution from diffraction spikes close to the afterglow position we model point spread functions for the appropriate bands using WebbPSF \citep{webbpsf}, and then scale and subtract these from star position. Photometry is measured in small (0.05 arcsec (blue) and 0.1 arcsec (red)) apertures and then corrected using tabulated encircled energy corrections. In addition to the direct photometry of the NIRCam images we also report a a K-band point based on folding the NIRSpec spectrum (see below), through a 2MASS, Ks filter. This both provides a better broadband SED and a direct comparison with ground based K-band observations. 
Details of photometric measurements are shown in Table~\ref{tab:photometry}. 

For NIRSpec, we utilise the available archive-processed level 3 two-dimensional spectrum (Figure~\ref{NIRSpec}). In this spectrum we clearly identify the trace of the optical counterpart, which appears effectively undetected until 2 microns and then rises rapidly. We also identify two likely emission lines which are offset from the burst position. These are consistent with the identification with H$\alpha$ and [O {\sc iii}] (4959/5007) at a redshift of $z=3.87$. Both of these lines lie within the F277W filter in NIRCam and support the identification of the nearby source as the origin of these lines. 

We extract the spectrum in two small (2 pixel) apertures. One of these is centred on the transient position, while the second is centred on the location of the emission lines. Since the offset between these two locations is only $\sim 0.2$ arcseconds there is naturally some contamination of each spectrum with light from both sources, but this is minimised by the use of small extraction apertures. The resulting 1D spectra are shown in Figure~\ref{NIRSpec}. The counterpart is very red, with a sharp break at 2 microns and an apparent emission feature at 2.15 microns. The spectrum then continues to rise to a possible second feature (or a change in the associated spectral slope) at around 4.5 microns.

\clearpage

\section{Supplementary Information}
\section{GRB 230307A in context}
\subsection{Prompt emission}
GRB 230307A is an exceptionally bright GRB. It has the second highest fluence of any GRB observed in more than 50 years of GRB observations \citep{burns23}. While it remains a factor of 50 less fluent than GRB 221009A, it is still a factor $\sim 2$ brighter than GRB 130427A, the third brightest burst. Bursts with these extreme fluences are rare. In Figure~\ref{fig:fluence_pl}, we plot the distribution of observed fluence for {\em Fermi}/GBM detected bursts. At the brighter end, the slope of the distribution is consistent with the expected $-3/2$ slope for a uniform distribution of sources. The extrapolation of this relation suggests that bursts like GRB 230307A should occur once every several decades. Notably, three bursts well above the extrapolation (GRB 130427A, GRB 230307A, GRB 221009A) may indicate that bright bursts arise more frequently than expected. However, observationally it is clear that GRB 230307A is, at least, a once-per-decade event.

The prompt light curve of GRB 230307A (Figure~\ref{fig:GBM}) shows two distinct emission features: an initial episode of hard emission from the trigger until $\approx 18$\,s, then a softer episode from $\approx 19$\,s onwards. These distinct episodes of hard and soft emission are strongly reminiscent of the long-duration merger GRB 211211A, but the initial pulse complex is $\sim 50$ per cent longer in GRB 230307A when compared to the $\sim 12$\,s duration seen in GRB 211211A \citep{rastinejad22}. The relative durations of the initial pulse complex in the two GRBs bear a striking resemblance to their relative time-averaged peak energies \citep[$936 \pm 3$\,keV vs $647 \pm 8$\,keV;][]{2023GCN.33411....1D, mangan21}. 
In GRB 211211A, substantial spectral evolution was seen to drive the light curve, and the underlying radiation mechanism was identified as fast-cooling synchrotron emission \citep{gompertz23}. The coherent development of the hardness ratio (Figure~\ref{fig:GBM}, lower) indicated similar spectral evolution in GRB 230307A, which the spectral analysis confirmed.
Indeed, as described in Section~\ref{gbm_analysis}, the time-resolved spectral analysis of the prompt emission revealed the presence of two spectral breaks in the GBM band, $E_\mathrm{break}$ and $E_\mathrm{peak}$, coherently becoming softer from 7.5 s up to 19.5 s. Also, in this case, the spectral indices indicate synchrotron emission in the marginally fast-cooling regime. From 19.5 s onwards (approximately when the softer and dimmer emission episode starts), the low-energy break $E_\mathrm{break}$ is continuously approaching the lower limit of the GBM band (8 keV),  presumably crossing it to enter the X-ray regime. Unfortunately, the lack of simultaneous observations in X-rays with another telescope, e.g. {\em Swift}/XRT, prevents us from fully tracing the evolution of the spectral break down to X-rays at later times, as was done for GRB 211211A.

The time-averaged {\em Fermi}/GBM spectrum of GRB 230307A across the $T_{90}$ interval is best fit with a cutoff power-law with $\alpha = 1.07 \pm 0.01$ and cutoff energy $936 \pm 3$\,keV \cite{2023GCN.33411....1D}. From this, we calculate a hardness ratio (the ratio of the 50 - 300\,keV photon flux to the 10 - 50\,keV photon flux) of $0.88^{+0.01}_{-0.02}$. This is higher than the value for 211211A ($0.57$) but comfortably within the $1\sigma$ distribution of hardness ratios for canonical long GRBs (i.e. with $T_{90} > 2$\,s) in the {\em Fermi} catalogue, which we calculate to be $0.66^{+0.51}_{-0.29}$ from the data in \citet{vonkienlin20}. Like GRB 211211A before it, GRB 230307A appears to have `typical' long GRB properties in terms of its time-averaged hardness ratio and its $T_{90}$. This strengthens the case for a significant number \citep{rastinejad22,troja22} of long-duration GRBs having been mistakenly identified as stellar collapse events. 

However, in some ways, GRB 230307A differs significantly from several of the other brightest GRBs. For example, the afterglow was relatively faint, while the burst was very bright. In Figure~\ref{fig:fluence_11hr}, we plot the prompt fluence in the 15-150 keV band against the X-ray afterglow brightness at 11 hours \cite[updated from][]{gehrels09,nysewander09}. The general trend between the afterglow brightness and fluence is seen; the best-fit slope to this relation is approximately one. So, while there is substantial scatter, there is a direct proportionality between the fluence and the afterglow brightness. Notably, while the afterglow and prompt emission of GRB 221009A were exceptionally bright (after correcting for the heavy foreground extinction), they were in keeping with this relatively broad relationship. GRB 230307A is different. Here we extrapolate the X-ray flux to 11 hours based on the measured X-ray flux at $\sim 1$ day and the decay slope. We also re-calculate the GRB 230307A fluence in the relevant 15-150 keV energy band for comparison to {\em Swift}/BAT. This burst is a notable outlier in the relation, with a faint X-ray flux for its extraordinary prompt brightness. The afterglow brightness depends both on the energy of the burst and the density of the interstellar medium; it is, therefore, possible that the location in this fluence -- afterglow brightness plane is indicative of a low-density medium, which would be consistent with expectations for such a large GRB - host offset.

It is also of interest that another burst in a similar location is GRB 211211A. This long burst has a clear signature of kilonova emission within its light curve. If GRB 230307A is a similar event, faint afterglows (relative to the prompt emission) may be an effective route for disentangling mergers from collapsars. 

To further compare the ratio between the X-ray brightness and the $\gamma$-ray fluence, we retrieve the X-ray light curve of all {\em Swift}-detected GRBs from the {\em Swift} Burst Analyser \citep{Evans2010a} and limit the sample to 985 long GRBs and 55 short GRBs with at least two XRT detections and measured BAT fluence. The fluences are taken from the {\em Swift}/BAT Gamma-Ray Burst Catalog\footnote{\href{https://swift.gsfc.nasa.gov/results/batgrbcat/index_tables.html}{https://swift.gsfc.nasa.gov/results/batgrbcat/index\_tables.html}} \citep{Lien2016a} and represent the measurements from 15 to 150 keV integrated over the total burst duration. We add to this sample the GRBs 170817A \citep[off-axis GRB;][]{Goldstein2017a} and 221009A \citep[brightest GRB detected to date;][]{burns23}. For the former, we retrieve the X-ray light curve from \citet{Hajela2022a} and use a $\gamma$-ray fluence of $2.4\times10^{-7}~\rm erg\,cm^{-2}$ \citep{Goldstein2017a}. For the latter, we take the X-ray light curve from the {\em Swift} Burst Analyser and assume a fluence of $0.007~\rm erg\,cm^{-2}$ (corrected from the 1-10000 keV fluence in \cite{burns23} to the 15-150 keV band). Following \cite{Schulze2014a}, we resample the X-ray light curves and normalise them by the $\gamma$-ray fluence on a grid defined by the observed $F_{\rm X} / \rm Fluence$ ratios and the time-span probed by the data. If no data are available at a specific time of the grid, we linearly interpolate between adjacent observations but do not extrapolate any data. Hence the paucity of observations at later times reflects the last time at which sources were detected by the {\em Swift}/XRT.

Short and long GRBs occupy the same part of the $F_{\rm X} / \rm Fluence$ vs time parameter space (Figure \ref{fig:GBM}). In contrast, GRB 230307A has an unprecedentedly low $F_{\rm X} / \rm Fluence$ ratio that is almost 10-fold lower than the faintest GRBs at the same time. To emphasise the uniqueness of GRB 230307A, we also show in the same figure the {\em Swift}/BAT-detected GRBs 050925, 051105A, 070209, 070810B, 100628A, 130313A, 170112A that evaded detection with {\em Swift}/XRT. The limits on their $F_{\rm X} / \rm Fluence$ ratio (shown by downward pointing triangles in that figure) are consistent with the observed range of $F_{\rm X} / \rm Fluence$ ratios, ruling out a selection bias against GRBs with lower than usual $F_{\rm X} / \rm Fluence$. Intriguingly, GRBs 080503, 191019A and 211211A had markedly low $F_{\rm X} / \rm Fluence$ ratios during the shallow decline phase of their X-ray light curves. Furthermore, GRB 211211A reached a value of $1.2\times10^{-9}~\rm s^{-1}$ at 120~ks, comparable to 
GRB 230307A.

\subsection{Counterpart Evolution} 
Although the afterglow of GRB 230307A was promptly detected thanks to TESS, this data was not available to the community for several days. Further follow-up was, therefore, much slower, and the counterpart was not discovered until the localisation was narrowed down to several sq. arcminutes, approximately 24 hours after the burst. The result is that the counterpart is poorly sampled (particularly in colour) during the early phases, while later observations suffer from typically modest signal-to-noise. 

The TESS observations detected a relatively bright (though not exceptional given the fluence of the burst) outburst, coincident with the prompt emission, likely peaking at $I<15$ \citep{fausnaugh23}. The afterglow was much fainter, apparently no brighter than $I=18$ in the minutes to hours after the burst was detected. It was relatively flat during this period, with a power-law through the first to last TESS observations decaying as $F(t) \propto t^{-0.2}$. The TESS and ground-based observations can be consistently modelled with a forward shock afterglow + kilonova (see Section~\ref{sec:6.1}).

There are no simultaneous colours at the time of the first ground-based afterglow detections (1.4 days), although extrapolation of the $r$-band detection with ULTRACAM to the WHITE detection with the {\em Swift}/UVOT suggests a relatively red colour (WHITE-r = 1.6 $\pm$ 0.4) . However, such an interpretation is difficult due both to the large photometric errors and the width of the WHITE filter on the {\em Swift}/UVOT.

Optical observations obtained multiple colours at an epoch $\sim 2.4$ days post-burst. These show the afterglow to have a blue colour with $g=22.35 \pm 0.26$, $i=21.68 \pm 0.09$ and $z=21.8$ \citep{2023GCN.33485....1G}. This is consistent with GRB afterglows in general(i.e $F_{\nu} \propto \nu^{-\beta}$ gives $\beta \approx 1$). Observations in the near infrared (NIR) were not undertaken until $\sim$ 10.4 days post-burst. However, these reveal a relatively bright K-band source. The inferred i-K(AB) $>$ 2.9 at this epoch is very red. Interpreted as a change in the spectral slope, it is $\beta \approx 2.5$. The K-band light hence appears to be in significant excess with respect to the afterglow expectations based both on optical data and on the X-ray light curve.

It is relevant to consider if such an excess could arise via extinction. However, this is not straightforward to explain. For a generic $\beta=1$ slope we expect $i-K(AB) \approx 1.1$. At $z=0$, to obtain $i-K(AB) = 2.9$ would correspond to a foreground extinction of $A_V \approx 4$. However, this would also predict $g-i \approx 3$, which is entirely inconsistent with the earlier observations. This problem becomes more acute for higher redshifts, where the bluer bands probe increasingly into the UV. 

The IR excess becomes extremely prominent by the time of the {\em JWST} observations. At 28.5 days, the source is detected in all bands but is very faint in the NIRCam blue channel (F070W, F115W, F150W) and rises rapidly (in $F_{\nu}$) through the redder bands (F277W, F356W, F444W). Expressed as a power-law, this is  $\beta \approx 3.1$ in the 2-5 micron region, and $\beta \approx 1$ between 0.7-1.5 microns. This does not match the expectations for any plausible spectral break in a GRB afterglow or any plausible extinction (where one would expect the slope to steepen towards the blue). This strongly implies that the red excess seen in the K-band at ten days and with JWST at 28.5 days is some additional component. Indeed, in the JWST observations, the other component, beginning at around 2 microns, is very clearly visible in both photometry and spectroscopy.

This component evolves exceptionally rapidly. In the K-band, the inferred decay rate from 11.5 to 28.5 days is $\sim t^{-3.5}$ expressed as a power-law or $\sim 0.25$ mag per day, if exponential. This is much faster than observed in GRB afterglows or 
supernovae. 
It is, however, consistent with the expectations for kilonovae. As shown in Figure 4, the overall evolution shows substantial similarity with AT2017gfo. To constrain the temporal and spectral evolution within a plausible physical model more accurately, we fit the multi-band photometry with afterglow and kilonova models. The outputs of these models are described in detail in section~\ref{sec:modeling}.


\subsection{Identification of the host galaxy}
\label{host_con}
Deep optical imaging of the field identifies several relatively bright galaxies in the vicinity of the sky position of GRB 230307A. Our preferred host galaxy is the brightest of these, which we denote as G1. It lies at $z=0.065$ and is offset 30 arcseconds (40 kpc in projection) from the location of the afterglow. Following the method of \cite{bloom02} this galaxy has a probability of chance alignment of $P_{\rm chance} \sim 0.09$ (see also \cite{2023GCN.33485....1G}). Although this is not extremely low, and so is only suggestive of a connection to the transient, 
we note that i) the luminosity of the late time counterpart at this redshift is very similar to AT2017gfo and ii) the spectral feature seen at 2.1 microns in AT2017gfo matches with the emission feature seen in the JWST spectroscopy of GRB 230307A. This is a broad line, but assuming they have the same physical origin, they fix the redshift to the range $0.04 < z < 0.08$. G1 is the only galaxy within this range in the field. The physical properties of this galaxy are outlined in section~\ref{gal_props}. 

Our MUSE observations provide redshifts for this galaxy and several others, also identifying a small group of galaxies (G2, G3, G4) at a common redshift of $z=0.263$.
All of these galaxies have $P_{\rm chance}$ values substantially greater than our preferred host. Furthermore, because of the larger redshifts, the implied offsets from GRB 230307A are $\gg100$kpc. This is larger than seen for any short GRB with a firmly identified host. We, therefore, disfavour these as plausible host galaxies for GRB 230307A. 

Deep JWST observations reveal no evidence of a directly underlying host galaxy for GRB 230307A, as would be expected if it had a collapsar origin. In particular, at late times, the faint source at the counterpart's location is consistent with a point-source (i.e. a subtraction of the PSF constructed by WebbPSF yields no significant residuals). However, we identify a faint galaxy, undetected in the blue and with F277W =$27.9 \pm 0.1$, offset only 0.3 arcseconds from the burst position. We designate this galaxy H1.

Our NIRSpec observations provide a redshift of $z=3.87$ for H1 based on the detection of [O III] (5007) and H$\alpha$. At this redshift, the offset is only $\sim 1.3$ kpc. Although many $z \sim 4$ galaxies are extremely compact \citep{straatman}, it seems likely that some stellar population from this galaxy does extend under the burst position, and there may be marginal evidence for extension in this direction in the F444W image. However, this region is neither UV-bright nor an emission line region where one may expect to observe massive stars. 

The galaxy photometry, performed in 0.1-arcsecond apertures and subsequently corrected for encircled energy assuming point-source curves is F070W$>$29.0, F115W=28.4 $\pm$ 0.3, F150W=28.6 $\pm$ 0.4, F277W=27.9 $\pm$ 0.1, F444W=28.3 $\pm$ 0.1, and the galaxy is only robustly detected in the redder bands (see Figure 2). We note that because of the proximity of the afterglow, we use a smaller aperture than may be optimal, although the galaxy is also compact. 
 
We can estimate the probability of chance alignment of this source with the GRB position via various routes. In principle, one can use number counts of galaxies on the sky in the multiple bands. These have recently been updated based on the first observations with {\em JWST} to provide number counts in appropriate bands \citep{windhorst23}. We find that $P_{\rm chance}$, following the approach of \cite{bloom02}
to be in the range $\sim$ 3-6 \% for F277W and F444W (with no bound in the filters where the galaxy is undetected). Alternatively, we also estimate the probability directly from the data. We extract sources within the field via Source Extractor to create a mask of objects within the field. In the brightest detection (F277W) approximately 5\% of the image is covered with objects of equal or brighter magnitude to H1, and we note that the burst position is {\em not contained} within this mask. This suggests that in this particular field, $P_{\rm chance} > 5\%$. 

The absolute magnitude of H1 is $M_i \sim -17.7$, and the H$\alpha$ star formation rate is approximately 1 M$_{\odot}$ yr$^{-1}$. The half-light radius of the galaxy is approximately 0.1 arcseconds (700 pc). Although limited information is available, these values are generally consistent with those of the long GRB population. The burst offset from its host galaxy is $\sim 2.5$ half-light radii. This is large but within the range seen for long-GRBs \cite{blanchard16}. 

In our X-shooter and MUSE observations there is no trace visible in 1D or 2D extractions at the source position, although a weak continuum is seen in the X-shooter spectrum when heavily binned. This is consistent with its faint magnitude at the time of the observations. At the location of Ly$\alpha$ at $z=3.87$ we place limits of $F < 2.5 \times 10^{-17}$ erg s$^{-1}$ cm$^{-2}$ assuming an unresolved line. 

We also examined both spectra for any emission lines at other redshifts. This is worthwhile given the strong emission lines often seen in long GRB hosts \citep{kruehler15}, which may make emission line redshifts possible, even if the host itself is undetected. However, despite deep observations, there are no visible emission lines consistent with no directly underlying host galaxy, consistent with a compact object merger, but not a collapsar. 

Unsurprisingly, there are also numerous faint galaxies in the JWST images. However, all of these have large $P_{\rm chance}$ values, and we do not consider them plausible host galaxies. 

Taken a face value, the probability of chance alignment for G1 (our preferred host) and H1 ($z=3.87$) is similar. However, the luminosity, lightcurve evolution and spectroscopic feature at the redshift of G1 offer strong support for it as the host galaxy of GRB 230307A. Furthermore, there is no straightforward, reasonably viable physical model that could explain the burst's extreme properties at $z=3.87$. This scenario would require extreme energetics, exceptionally rapid evolution and yields unphysical outcomes in standard GRB or supernovae scenarios. We outline this in detail in section~\ref{hiz}. 

\subsection{Host galaxy properties}
\label{gal_props}
To better understand the properties of G1, the likely GRB host galaxy, we performed a fit to both the MUSE spectrum and photometric measurements from the far-UV to the mid-IR. For the photometric measurements, we retrieved science-ready coadded images from the \textit{Galaxy Evolution Explorer} (GALEX) general release 6/7 \cite{Martin2005a}, DESI Legacy Imaging Surveys (LS; \cite{Dey2018a}) data release 9, and re-processed WISE images \cite{Wright2010a} from the unWISE archive \cite{Lang2014a}\footnote{\href{http://unwise.me}{http://unwise.me}}. The unWISE images are based on the public WISE data and include images from the ongoing NEOWISE-Reactivation mission R7 \cite{Mainzer2014a, Meisner2017a}. We measured the brightness of the galaxy G1 using the Lambda Adaptive Multi-Band Deblending Algorithm in R (\textit{LAMBDAR} \cite{Wright2016a}) and the methods described in \cite{Schulze2021a}. We augment the SED with \textit{Swift}/UVOT photometry in the $u$ band and our 6-band JWST/NIRCAM photometry. The photometry on the UVOT images was done with \textit{uvotsource} in \textit{HEASoft} and an aperture encircling the entire galaxy. For JWST photometry, we used a 6-arcsec circular aperture, which allows us to gather all the observed light observed in JWST filters from the host galaxy. All measurements are summarised in Table \ref{tab:hostmag}. 

To derive the main physical properties of the host galaxy, such as its stellar mass, we employ two separate methodologies based on the photometric and spectroscopic data available for the host, and finally compare the results to assess the robustness of our conclusions. We first fit the multi-wavelength (0.1--4.4 $\mu$m) dataset using the \texttt{prospector} python package \cite{Johnson2021}, which allows us to model the host galaxy spectrum starting from its main constituents, namely a set of stellar population base spectra, built from the Flexible Stellar Population Synthesis (FSPS) package \cite{Conroy10}, and combined with a specific star-formation history (SFH) model. Moreover, we have also considered a fixed attenuation model based on the Calzetti \cite{Calzetti2000} attenuation curve, and an additional nebular model originating from the gas component, which is built using the \texttt{Cloudy} photo-ionization code \cite{Ferland2017}, and considering the FSPS stellar population as ionising sources. We have adopted a parametric SFH model, which is described by a delayed-exponential model where the star-formation rate varies as a function of time $t = t_\mathrm{age} - t_\mathrm{lt}$, with $t_\mathrm{lt}$ being the lookback time \cite{Johnson2021}, as $\mbox{SFR} \propto (t/\tau) \exp(-t/\tau)$, with $\tau$ being the $e$-folding time. We finally have used the \texttt{dynesty} \citet{Speagle2020a} ensemble sampler to reconstruct the posterior distribution.  

The results of the prospector analysis are shown in Fig. \ref{fig:prospector}. We obtain a mass value of the living stars of $M_{*} = 2.37 (+0.24,-0.35) \times 10^9$ M$_{\odot}$ yr$^{-1}$. The mass of all stars ever formed is $0.20 (+0.02,-0.04)$ dex larger. The light-weighted stellar age resulting from the fit is $1.13 (+1.49,-0.36)$ Gyr.

An alternative to parametric SED fits is to use synthetic stellar population SEDs as templates and combine them to fit the galactic spectra (the underlying assumption being instantaneous star formations rather than continuous functions of time).
We can use the spectral synthesis from the BPASSv2.2.2\cite{bpass2017, bpass2018} binary populations and create templates with {\tt hoki}\cite{hoki} that are compatible with the {\tt ppxf} fitting package\cite{ppxf}, as described in \cite{stevance23}.   
Because SED fitting has a high level of degeneracies (see \cite{ppxf}), at first we do not fit all 13 BPASS metallicities at once with {\tt ppxf}, as this can result in unphysical results (see discussion in \citealt{stevance23}); instead we fit the metallicities individually to find which ones result in the best fits on their own. 
We find that a low Z (0.001) population and solar metallicity population (Z=0.014) result in decent fits, but the low metallicity population fails to predict a young stellar component that is seen in the images, whilst the solar metallicity fit fails to accurately match the H$\beta$ and neighbouring absorption features in the blue part of the spectrum.
So we then fit the galaxy simultaneously with Z=0.001 and Z=0.014 templates, and retrieve a good fit shown in Figure \ref{fig:BPASS_fit_sfh} alongside the recovered SFH.

We find evidence of three main stellar populations: $>$95\% of the mass is found in lower metallicity (Z=0.001) stars with ages ranging from a few Gyr to 10 Gyr, with a peak of star formation around 5 Gyr; $>$4.7\% of the mass originates from a solar metallicity population (Z=0.014) that formed around 400 Myr ago; finally a small fraction ($<$0.05\%) of the stellar mass in the host originates from the star-forming regions with ages a few Myrs.

The details of the age distributions and exact metallicity values can be model dependent so we also fit the integrated galaxy spectrum with the single stellar population synthesis code STARLIGHT \cite{Cid2005}, which uses stellar populations based on 25 different ages and six metallicity values \cite{Bruzual03}, and a Chabrier IMF \cite{chabrier03}.
The SFH retrieved by this method is more complex and would require odd configurations (including some high metallicties at old ages and low metallicities around 100 Myr, which is counter-intuitive, unless inflow from pristine gas will trigger a burst of SFR), but it also finds that overall the galaxy is dominated by an old population with lower metallicity and has a younger component at higher metallicity. 
In Figure \ref{fig:BPASS_fit_sfh} we show a comparison of the STARLIGHT and BPASS fits in the bottom left panel and see that they are very similar, despite STARLIGHT containing 6 different metallicities and assuming solely single star populations. 
This highlights the level of degeneracy we face when performing galaxy SED fits.
We leave further comparisons to a follow-up study dedicated to the host and the progenitor populations of GRB 230307A, where we will also present detailed, specially resolved, fits to the datacube including its kinematics.

For now we use the BPASS integrated fits to infer the stellar mass and the star formation rate of the host of GRB 230307A, as the fit and SFH is more convincing that the one obtained with STARLIGHT.
We find that there are currently $M_{*}=1.65\times10^{9} M_{\odot}$ of living stars (corresponding to 3.1 $\times10^{9} M_{\odot}$ at ZAMS) in G1. 
Using the nebular component retrieved from subtracting the fit of the stellar component to the observed data, we can also estimate the star formation rate and metallicity. 
From the H$\alpha$ feature we estimate that the SFR is 5.47$\pm 0.30 \times 10^{-1}\,M_{\odot} {\rm yr}^{-1}$ using the Kennicutt formulation \cite{Kennicutt98}, and using the N2 index, in the CALIFA formulation \cite{Marino13}, we infer an oxygen abundance of 8.20 $\pm$ 0.16 (12 + log(O/H)). 

There are qualitative similarities between the host of GRB 230307A and NGC 4993\cite{stevance23}, the host of the first confirmed kilonova (they are both dominated by an older stellar populations and include a younger more metal rich component), but there are some key differences: NGC 4993 was a lenticular galaxy without a clear young component, whereas the host of GRB 230307A shows clear spiral arms and star forming regions. 
Another major difference is that the metallicity of the old population in this galaxy is 10 times lower than that of NGC 4993 (Z=0.001 compared to Z=0.010), which will influence the stellar evolution of potential progenitors. Finally, NGC 4993 had a large stellar (and presumably dark halo) mass $M_* \approx 10^{11}$ M$_{\odot}$ \cite{stevance23}, a factor of $>50$ larger than the host of GRB 230307A. 

The location of GRB 230307A relative to its host galaxy is consistent with these properties. In particular, the low mass of the galaxy suggests a modest gravitational potential such that binaries with velocities of a few hundred km s$^{-1}$ can readily escape. The large offset also suggests that the binary is formed from the older stellar population. 





\subsection{Properties of the brightest GRBs}
GRB 230307A is the second brightest\footnote{We use brightest here as an indicator of the total fluence in the prompt emission}
burst observed in over 50 years of observations \citep{burns23}. If it arises from a compact object merger, this implies that such bright bursts can be created in mergers. Indeed, such a picture appears likely based on GRB 211211A \citep{rastinejad22,troja22,yang22}, the sixth brightest burst. Of the ten brightest bursts observed by the {\em Fermi}/GBM, and subsequently localised at the arcsecond level, three have apparently secure associations with supernovae (GRB 130427A, GRB 171010A, GRB 190114C), and two (GRB 211211A, GRB 230307A) are associated with kilonovae, and hence mergers. Of the remaining five, one lies at $z=1.4$ and has energetics which suggest a collapsar; three have no redshift information, although one of these (GRB 160821A) lies in proximity to several galaxies at $z=0.19$; and one is GRB 221009A whose associated with a supernova remains unclear \citep{2023arXiv230111170F,Shrestha2023arXiv,levan2023}, although recent observations suggest a collapsar with 
an associated supernova is most likely \citep{blanchard23}. Within this very bright population, collapsars are likely as common as mergers. 

\section{Event rates}
One key question of interest is the likely event rate for such merger GRBs. A simple estimate of the event rate associated to a single event is given by 
\begin{equation}
R = \frac{1}{\Omega t V_\mathrm{max}}.
\end{equation}
Here $\Omega$ reflects the fraction of the sky covered by the detection mission, $t$ the effective mission duration (accounting for the duty cycle) and $V_\mathrm{max}$ the maximum co-moving volume within which a burst with the same properties could be identified. 

\begin{figure}
   \centering
   \includegraphics[width=14cm]{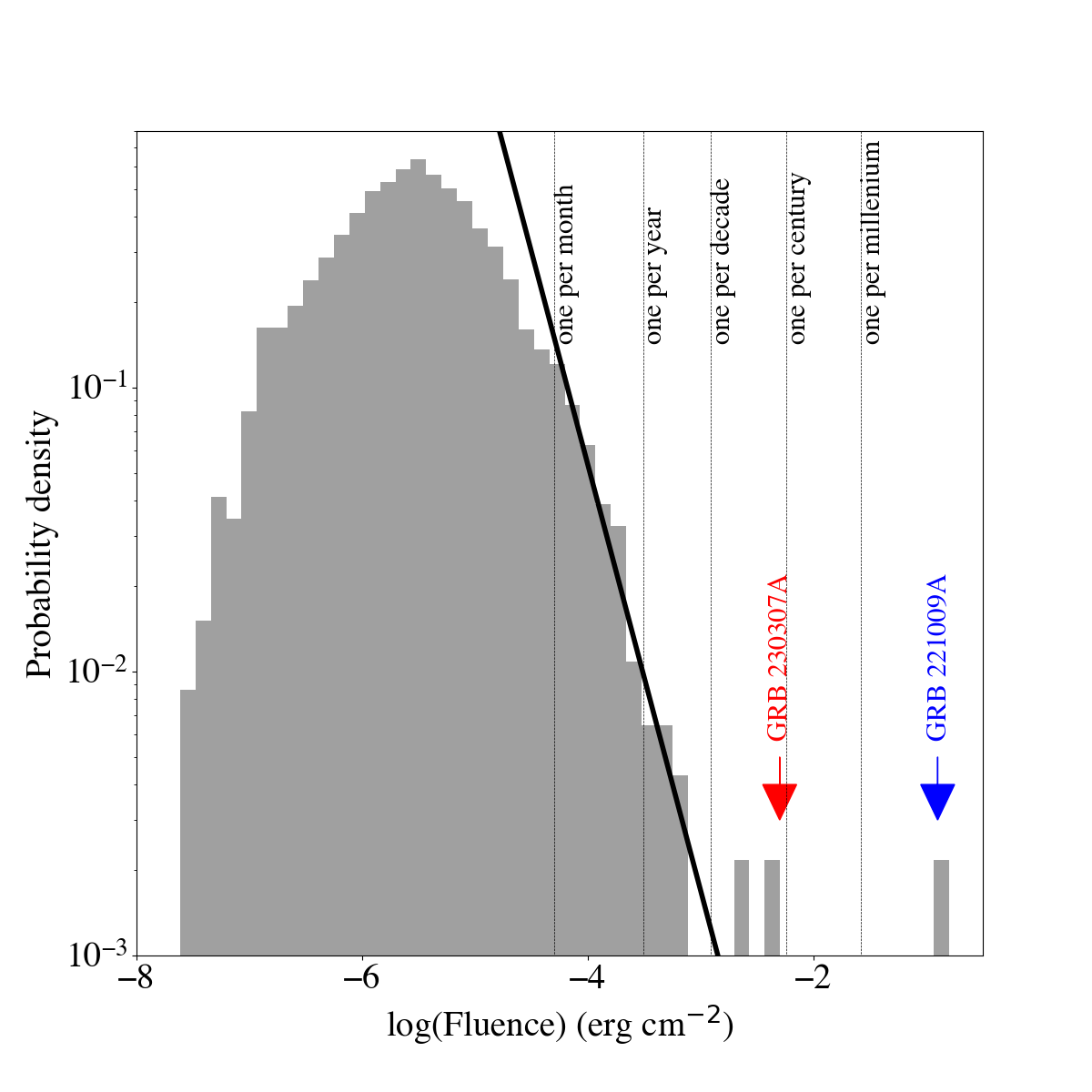}
      \caption{The distribution of measure fluence from the {\em Fermi}/GBM catalogue \cite{vonkienlin20}. The solid line shows an expected slope of $-3/2$ for a uniform distribution. The faint end deviates from this line because of incompleteness. At the brighter end, there are three bursts which appear to be extremely rare, GRB 130427A, GRB 221009A \citep{burns23} and GRB 230307A. To indicate the apparent rarity we also plot lines representing the expected frequency of events under the assumption of a $-3/2$ slope. We would expect to observe bursts akin to GRB 230307A only once per several decades.   }\label{fig:fluence_pl}
   \end{figure}

\begin{figure}
   \centering
   \includegraphics[width=\hsize]{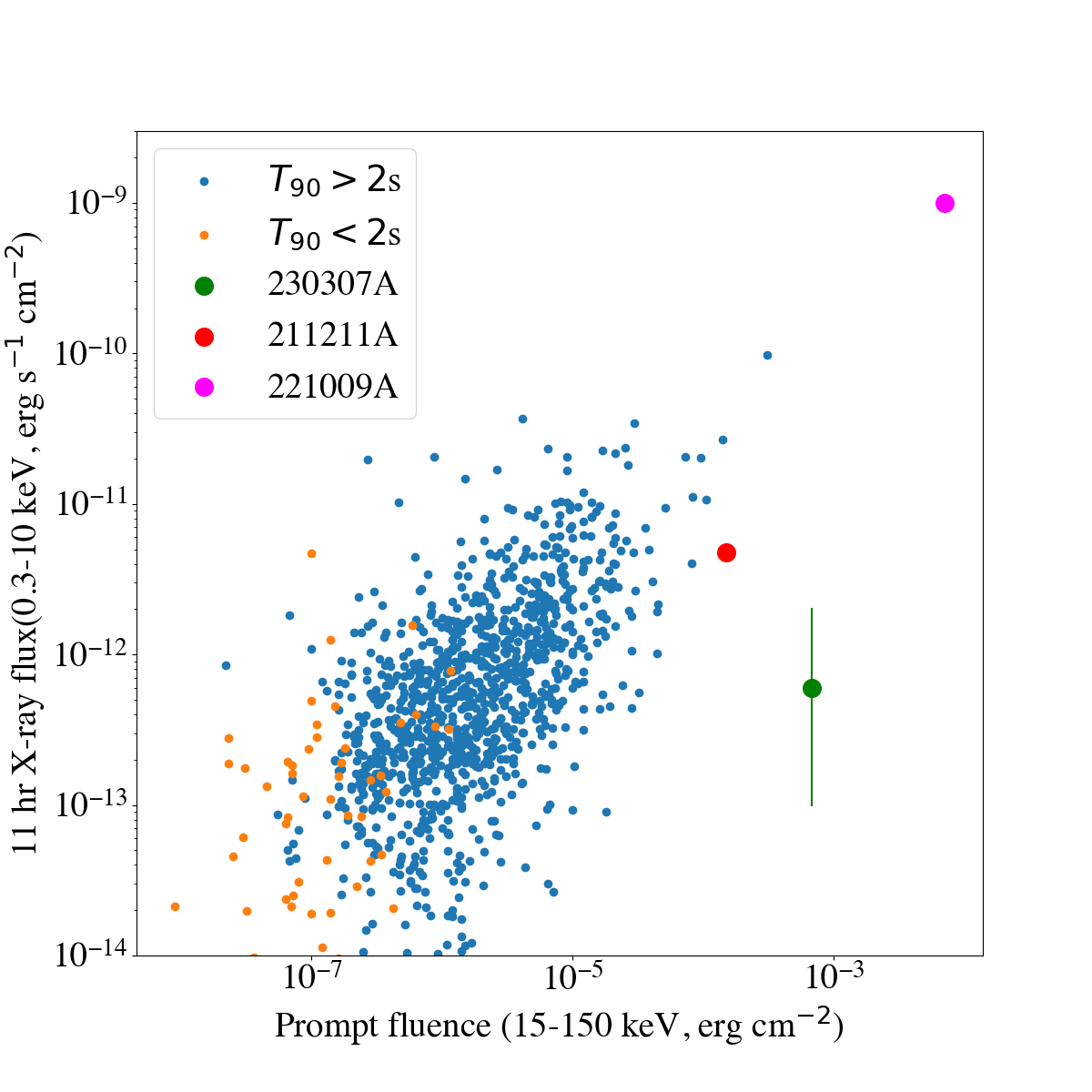}
      \caption{A comparison between the prompt fluence (in the 15-150 keV band) and the X-ray flux at 11 hours for {\em Swift} GRBs with GRB 221009A and GRB 230307A added, updated from \cite{gehrels09,nysewander09}. The general 1:1 trend between the prompt fluence and  X-ray brightness can clearly be seen, although it has a significant scatter, although a very rare event, GRB 221009A apparently lies on the same relation. However, GRB 211211A and GRB 230307A are clearly outliers to this relation with GRB 230307A occupying a region devoid of other GRBs. This very faint afterglow compared to the prompt emission may be related to a location at large projected offset from its host galaxy in which the density of the ambient medium is very low. 
}\label{fig:fluence_11hr}
   \end{figure}

\begin{figure}
   \centering
   \includegraphics[width=\hsize]{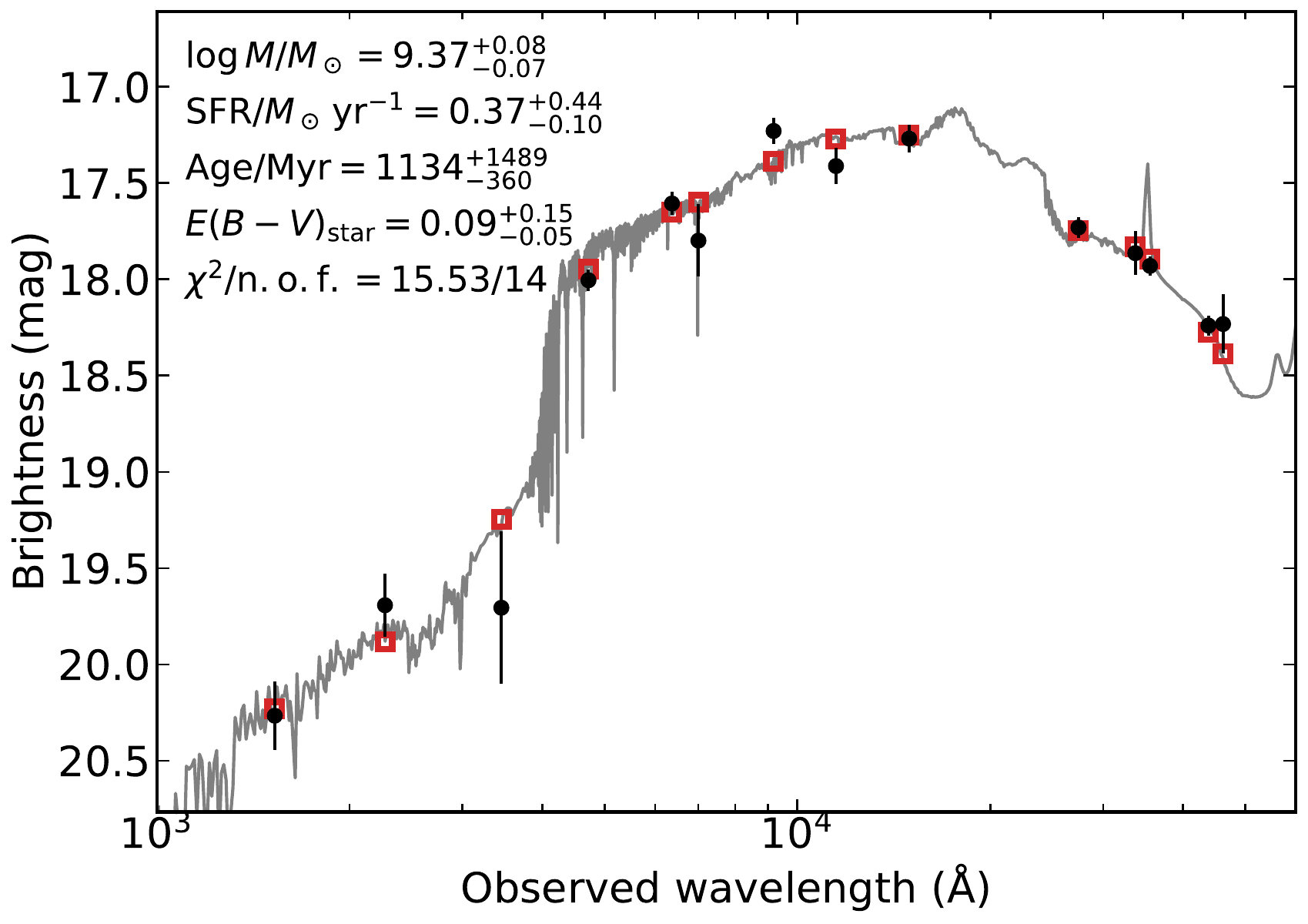}
      \caption{Spectral energy distribution (SED) of the galaxy G1 from 1000 to 60,000~\AA\ (black data points) and its best fit with the \textit{prospector} SED fitting code (grey shaded curve). In the top right, we also report the values of the model parameters and their $1\sigma$ uncertainties. The red squares represent the model-predicted magnitudes. The fitting parameters are shown in the upper-left corner. The abbreviation `n.o.f.' stands for the number of filters.
      }\label{fig:prospector}
   \end{figure}

\begin{figure}
   \centering
   \includegraphics[width=\hsize]{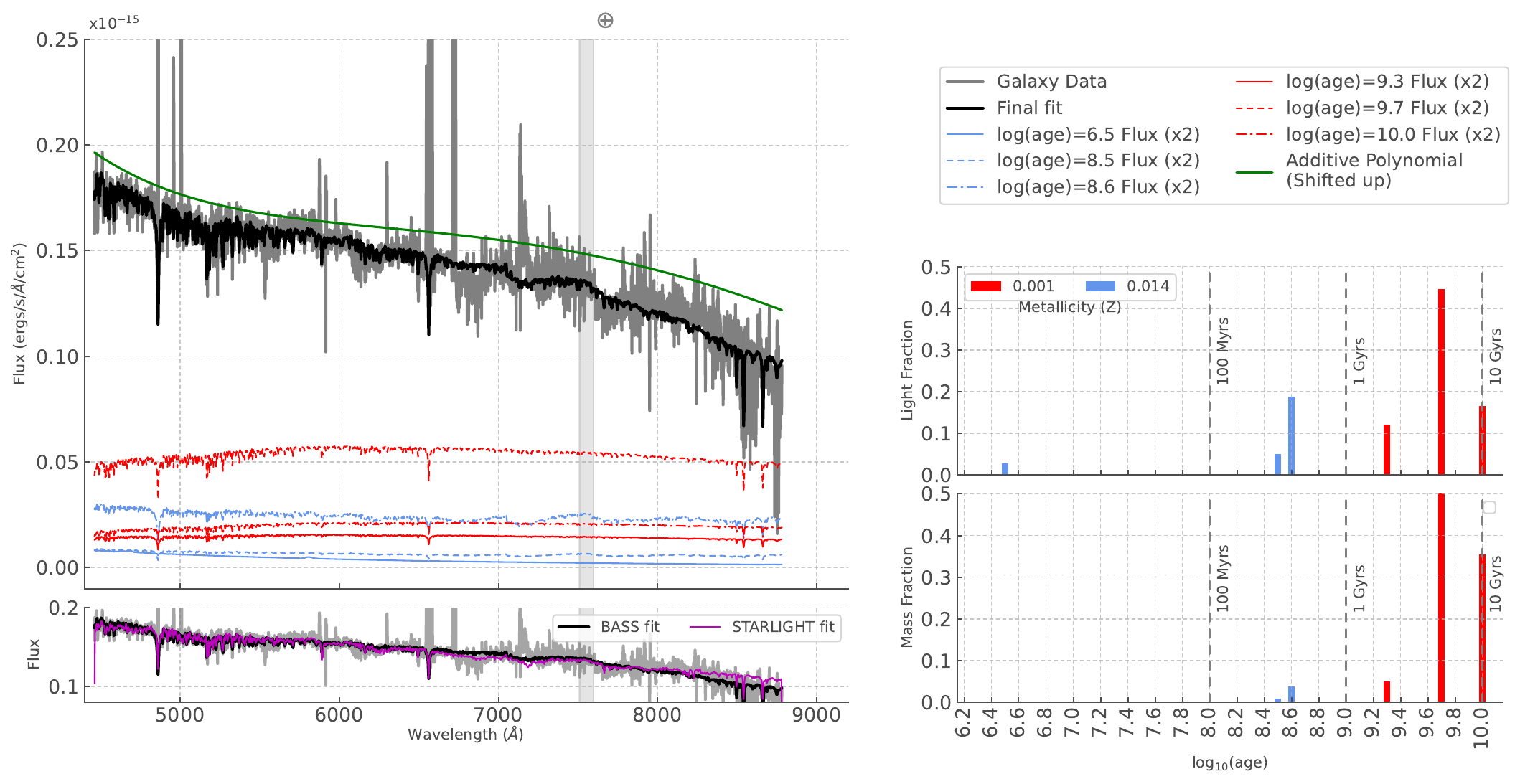}
      \caption{SED fit (top left panel) and Star Formation History (right hand panel of the host galaxy obtained with {\sc BPASSv.2.2.2}--{\tt hoki} templates fit with {\tt ppxf}. We also include a comparison of the final fits obtained with BPASS (2 metallicities) and STALRIGHT (6 metallicites) in the bottom left panel to highlight how similar both fits are. }\label{fig:BPASS_fit_sfh}
   \end{figure}

For GRB 230307A, $\Omega=0.65$ (average for the {\em Fermi}/GBM) and $t \approx 15$ years. $V_\mathrm{max}$ is more complicated: as shown in Figure~\ref{fig:fluence_pl}, the fluence distribution for GBM bursts extends to $\sim 10^{-8}$ erg cm$^{-2}$ and is likely complete to around $10^{-6}$ erg cm$^{-2}$. Given the extreme brightness of GRB 230307A, it would likely have been recovered to a distance $\sim 50$ times greater than its observed distance. If at $z=0.065$ the inferred $z_\mathrm{max}=2.03$ or $V_\mathrm{max} = 630$ Gpc$^{-3}$. In this case, the inferred rate of such bursts becomes extremely small, $R \approx 1.6 \times 10^{-4}$ Gpc$^{-3}$ yr$^{-1}$. However, in practice, such bursts would not readily be identified at such redshifts since neither supernova nor kilonova signatures could be observed. A more realistic estimate would correspond to the distance at which associated supernovae can be either identified or ruled out with moderate confidence. In this case 
$z_\mathrm{max} = 0.5$ (also adopted by 
\cite{yin23}), $V_\mathrm{max} = 29$ Gpc$^{3}$, and  
$R \approx 3.5 \times 10^{-3}$ Gpc$^{-3}$ yr$^{-1}$. 

These rate estimates also assume that GRB 230307A is the {\em only} merger-GRB to have occurred within the 15-year lifetime of the {\em Fermi}/GBM. This is almost certainly not the case. Indeed, GRB 211211A was also identified by {\em Fermi}/GBM and has rather similar estimates of the intrinisc rate \citep[$5.7 \times 10^{-3}$ Gpc$^{-3}$ yr$^{-1}$,][]{yin23}. 

However, even the interpretation of $\sim 2$ events is problematic. In particular, the $V/V_\mathrm{max}$ for GRB 230307A is $0.004$, and for GRB 211211A $=0.005$ (again assuming $z_\mathrm{max} = 0.5$). For a sample average of uniformly distributed sources of comparable energy or luminosity, we expect $V/V_\mathrm{max} \sim 0.5$). That the initial identification of such a population should arise from bursts with such extreme $V/V_\mathrm{max}$ values is surprising, but may reflect that these bursts are the brightest, which likely encouraged a detailed follow-up. However, it is improbable that they represent the only such bursts observed, and we should expect a much larger population. 

To better quantify this, we extend our analysis to the {\em Swift} bursts and utilise the fluence of GRB 230307A converted to a 15--150 keV equivalent fluence using the observed spectral parameters. At $z=0.065$, $E_\mathrm{iso} \mbox{(15--150)~keV} \sim 7 \times 10^{51}$ erg, and for GRB 211211A $E_\mathrm{iso} \mbox{(15-150)~keV} = 2 \times 10^{51}$ erg. As expected, low energy events dominate the low redshift GRB population. However, at $z<0.5$, there are 12 (out of 42) bursts with $E_\mathrm{iso} \gtrsim 10^{51}$ erg. This includes some further supernova-less GRBs, in particular GRB 060614 ($E_\mathrm{iso} = 9 \times 10^{50}$ erg), GRB 191019A ($E_\mathrm{iso} = 2.0 \times 10^{51}$ erg), and some bursts for which supernova searches have not been reported (e.g.\ GRB 150727A, GRB 061021, and the `ultra-long' GRB 130925A). This sets an upper limit on the number of bursts at low redshift, which may be associated with mergers. In practice, selection effects would support a scenario where mergers generate a larger fraction of these bursts. In particular, the afterglows of GRB 230307A and GRB 211211A appear to be faint, despite the bright prompt emission. Such afterglows are difficult to find and may evade detection. In these cases redshifts may only be obtained from host galaxies. The associations may not be obvious if the bursts are offset from host galaxies at moderate redshifts. Such follow-up may occur late after the burst, or optical afterglow non-detections may lead to a lack of optical/IR follow-up because of uncertainty regarding the optical brightness of the event or suggestions it may be optically dark because of host galaxy extinction. Finally, given the afterglow brightness issues, it is possible that the small fraction of bright GRBs without redshift measurements may arise from a similar channel. These observations would imply that between 30-70\% bursts at $z<0.5$ and $E_\mathrm{iso} \gtrsim 10^{51}$ erg could arise from mergers, although it is likely less. A modest number of events at higher redshift is consistent with the observations, and would alleviate concerns regarding $V/V_{max}$ for GRB 211211A and GRB 230307A.

This fraction is surprisingly high given the strong evidence that long GRBs arise from broad-lined type Ic supernovae and short GRBs from compact object mergers. However, the dominant contributors to the long-GRB supernova connection occur at low energy, and belong to a population of low luminosity GRBs (LLGRBs) \cite{liang07}. In a significant number of these, we may observe a energy source in the prompt emission separate from the highly relativistic jet seen in on-axis, energetic bursts. For example, the long-lived, soft nature of some bursts suggests a contribution from shock breakout or cocoon emission. If, for this reason, the luminosity function of collapsar GRBs is steeper at low luminosity than that of merger-GRBs, it is possible that at low luminosity the long GRB population is dominated by collapsars, while at high luminosity the contribution of mergers is significant. Such an interpretation is not without problems, given the star-forming nature of long-GRB hosts and their typically small offsets from their host galaxies. However, it is a logical investigation for future work.

\section{Modelling}
\label{sec:modeling}
\subsection{Light curve modelling}\label{sec:6.1}

In order to shed light onto the properties of the jet and, even more importantly, to separate the contribution of the kilonova from that of the jet afterglow in the UVOIR bands, we modelled the multi-wavelength light curves from radio to X-rays as a superposition of synchrotron emission from the forward shock driven by the jet into the interstellar medium (ISM), following \cite{Salafia2022,mei22}, and blackbody emission from the photophere of a kilonova, using the simple single-component model of \cite{hotokezaka20}.

The forward shock synchrotron emission model has eight parameters, namely the isotropic-equivalent kinetic energy in the jet $E_\mathrm{K}$, its initial bulk Lorentz factor $\Gamma_0$, its half-opening angle $\theta_\mathrm{j}$, the ISM number density $n$, the fraction $\xi_\mathrm{N}$ of ISM electrons that undergo diffusive shock acceleration in the forward shock, the fraction $\epsilon_\mathrm{e}$ of the shock downstream internal energy that is shared by such electrons, the slope $p$ of the power law $\mathrm{d}N_\mathrm{e}/\mathrm{d}\gamma\propto\gamma^{-p}$ that describes the Lorentz factor (as measured in the shock downstream comoving frame) distribution of the accelerated electrons as they leave the acceleration region, and the fraction $\epsilon_\mathrm{B}$ of the shock downstream internal energy that is shared by a small-scale, turbulence-driven, random magnetic field. The shock hydrodynamics is computed from energy conservation and accounts for the lateral expansion of the shock \cite{Salafia2022}. The effective electron energy distribution is computed accounting for the cooling induced by synchrotron and synchrotron-self-Compton emission, including an approximate treatment of the Klein-Nishina suppression of the Thomson cross section \cite{Salafia2022}. In computing flux densities, the synchrotron surface brightness of the shock is integrated over equal-arrival-time surfaces to account for the effects of relativistic aberration and latitude-dependent retarded times on the spectral shape \cite{Panaitescu2000}. 

The kilonova model \cite{hotokezaka20} assumes spherical ejecta expanding homologously, $v=r/t$, and featuring a power law density profile $\rho(r,t) \propto t^{-3}v^{-\delta}$ between a minimum and a maximum velocity, $v_\mathrm{ej}\leq v \leq v_\mathrm{ej,max}$. The density normalization is set by the total ejecta mass $M_\mathrm{ej}$. In general, the model allows for the ejecta opacity (assumed grey) $\kappa$ to be piecewise-constant within the profile, but here we assume a uniform opacity across the ejecta for simplicity. The model divides the ejecta into 100 small shells and computes the heating rate and thermalization efficiency within each. This allows for the derivation of the internal energy evolution in each shell and eventually the computation of the photospheric luminosity $L_\mathrm{KN}$ in the diffusion approximation. The fixed ejecta opacity also allows for the computation of the optical depth and hence for the identification of a photospheric radius, which then sets the effective temperature $T_\mathrm{KN}$ by the Stefan-Boltzmann law. In our modelling of GRB~230307A, we computed the flux density by simply assuming pure blackbody emission with the given luminosity and effective temperature at each given time.
We fixed $v_\mathrm{max}=0.6 c$ and left $M_\mathrm{ej}$, $v_\mathrm{ej}$, $\kappa$ and $\delta$ as free parameters.

To carry out the model fitting, we defined an asymmetric Gaussian log-likelihood term for the $i$-th datapoint, which corresponds to an observation at time $t_i$ and in a band whose central frequency is $\nu_i$, as
\begin{equation}
    \ln \mathcal{L}_i = -\frac{1}{2}\frac{(F_{\mathrm{\nu,m}}(\nu_i,t_i)-F_{\nu,\mathrm{obs,i}})^2}{\sigma_i^2 + f_\mathrm{sys}^2F_{\mathrm{\nu,m}}^2} 
     -\ln{\left[\sqrt{2\pi\left(\sigma_{\mathrm{l},i}^2+f_\mathrm{sys}^2F_{\mathrm{\nu,m}}^2\right)}+\sqrt{2\pi\left(\sigma_{\mathrm{h},i}^2+f_\mathrm{sys}^2F_{\mathrm{\nu,m}}^2\right)}\right]},
\end{equation}
where $F_{\nu,\mathrm{m}}(\nu,t)$ is the flux density predicted by the model, $F_{\nu,\mathrm{obs},i}$ is the measured flux density, the one-sigma error reflects the potentially asymmetric error bars
\begin{equation}
    \sigma_i = \left\lbrace\begin{array}{lr}
    \sigma_{\mathrm{l},i} & \mathrm{if\,} F_{\nu,\mathrm{m}}(\nu_i,t_i)\leq F_{\nu,\mathrm{obs},i}\\
    \sigma_{\mathrm{h},i} & \mathrm{if\,} F_{\nu,\mathrm{m}}(\nu_i,t_i)> F_{\nu,\mathrm{obs},i}\\
    \end{array}\right.,
\end{equation}
and we introduced a fractional systematic error contribution $f_\mathrm{sys}$, which we take as an additional nuisance parameter, to account for potential inter-calibration uncertainties between different instruments and for the fact that error bars typically only account for statistical uncertainties. For X-ray detections, we fit the integrated flux and the spectral index independently, with an analogous term for each (but with no systematic error contribution for the spectral index). Upper limits were treated simply by setting $F_{\nu,\mathrm{obs},i}$ equal to the reported upper limit, $\sigma_{\mathrm{h},i}=F_{\nu,\mathrm{obs},i}/10$ and $\sigma_{\mathrm{l},i}=10 F_{\nu,\mathrm{obs},i}$. The final log-likelihood was taken as the sum of these terms. 

\begin{figure}
    \centering
    \includegraphics[width=\hsize]{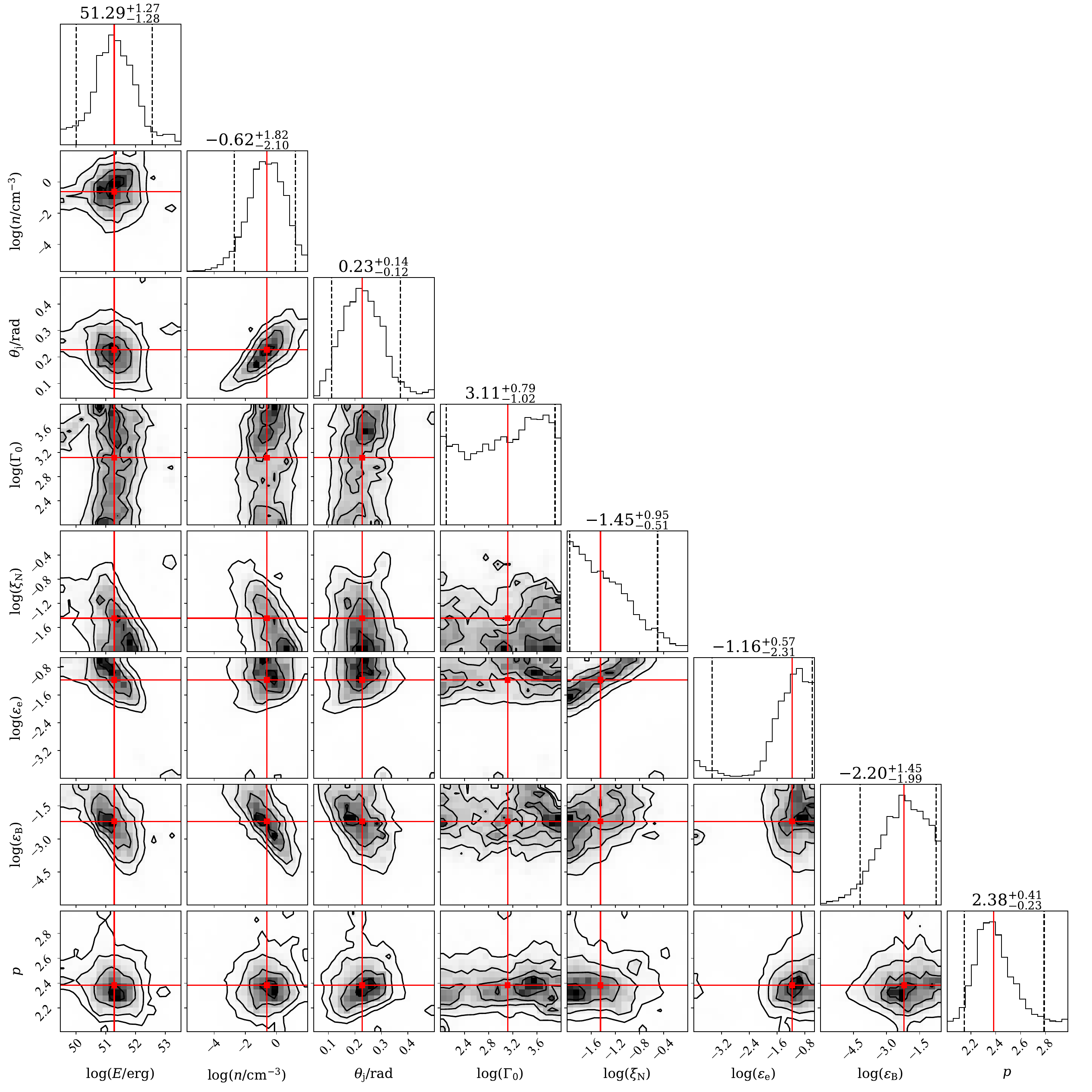}
    \caption{Corner plot of posterior probability density from multi-wavelength light curve fitting, limited to the relativistic jet afterglow parameter space. Histograms on the diagonal show the marginalized posterior probability densities on the parameters constructed from our MCMC posterior samples. Dashed black lines show the 90\% credible interval, while red lines show the medians. The remaining plots show the one, two and three sigma equivalent contours of the joint posterior probability densities of parameter pairs. Red lines and squares mark the medians.}
    \label{fig:light_curve_corner_plot_AG}
\end{figure}

\begin{figure}
    \centering
    \includegraphics[width=\hsize]{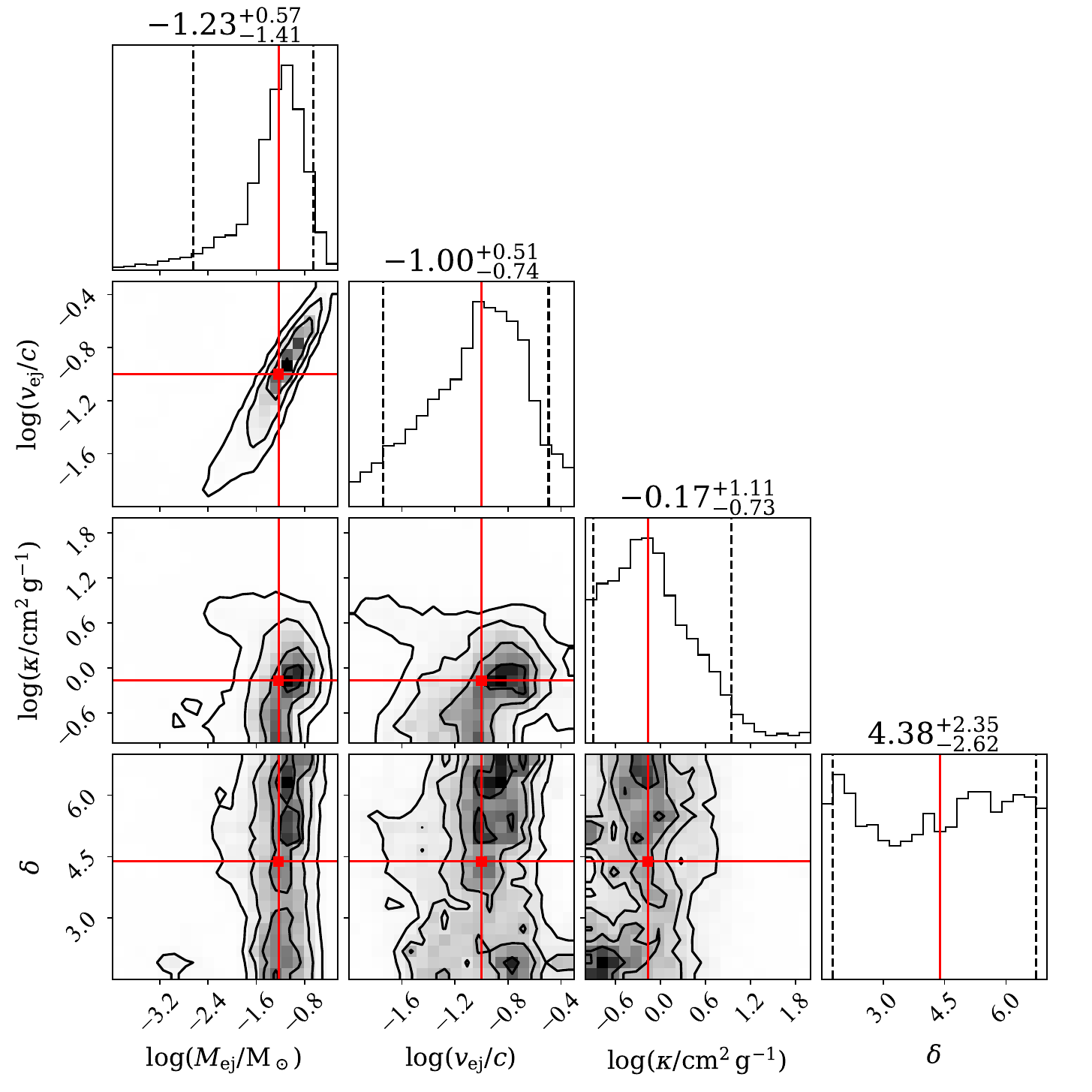}
    \caption{Corner plot of posterior probability density from multi-wavelength light curve fitting, limited to the kilonova parameter space. Similar to figure \ref{fig:light_curve_corner_plot_AG}, but for the kilonova model parameters.}
    \label{fig:light_curve_corner_plot_KN}
\end{figure}

\begin{figure}
    \centering
    \includegraphics[width=\hsize]{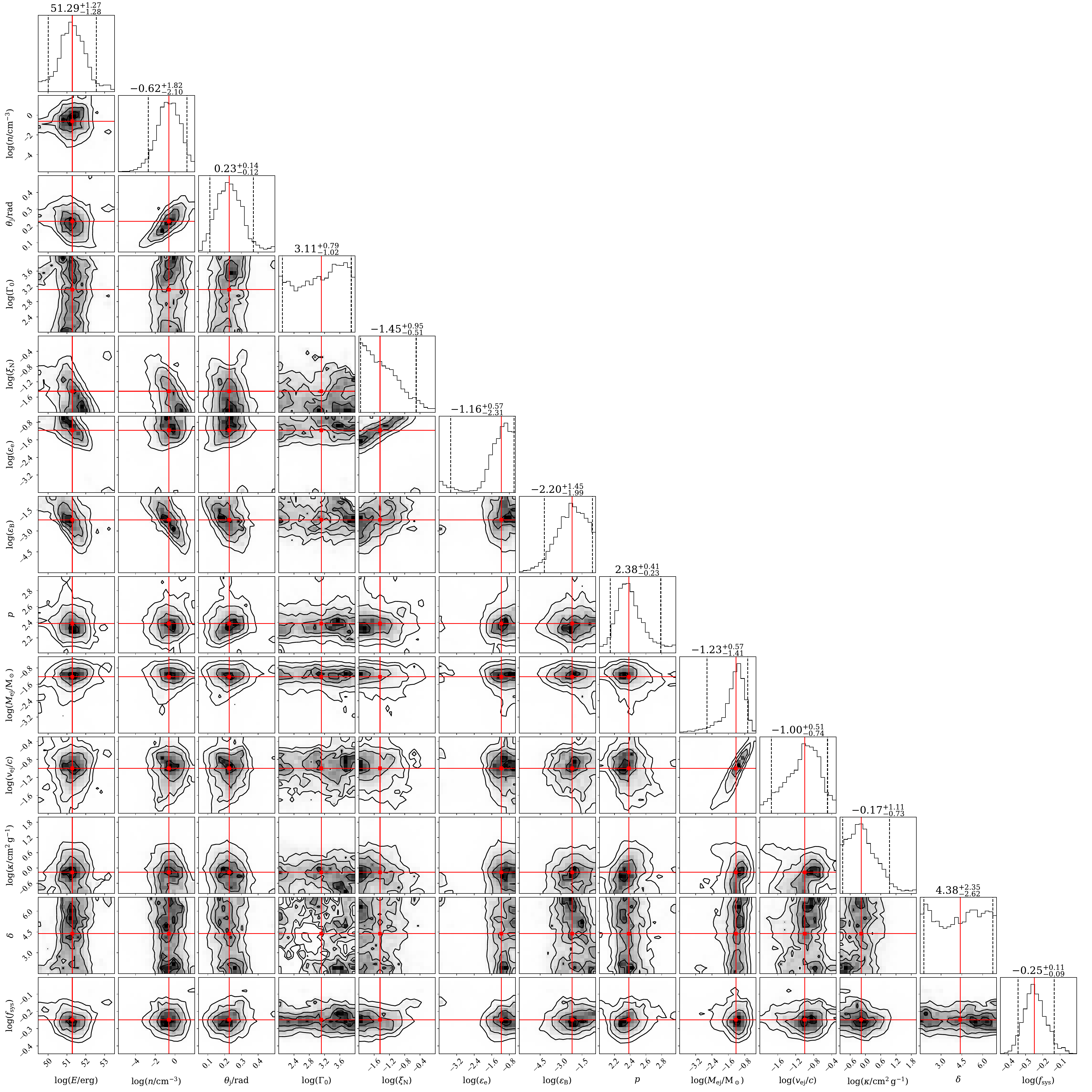}
    \caption{Corner plot of posterior probability density from multi-wavelength light curve fitting. Similar to figures \ref{fig:light_curve_corner_plot_AG} and \ref{fig:light_curve_corner_plot_KN}, but showing all model parameters, including the nuisance parameter $f_\mathrm{sys}$.}
    \label{fig:light_curve_corner_plot_full}
\end{figure}

In order to derive a posterior probability density on our 13-dimensional parameter space, we assumed the priors reported in Table \ref{tab:light_curve_mcmc} and we sampled the posterior with a Markov Chain Monte Carlo approach using the \texttt{emcee} python module \cite{emcee}, which implements the Goodman and Weare \cite{Goodman2010} affine-invariant ensemple sampler. The medians and 90\% credible intervals of the marginalised posteriors on each parameter obtained in this way are reported in Table \ref{tab:light_curve_mcmc}. The posterior is visualised by means of corner plots in Figures \ref{fig:light_curve_corner_plot_AG} (jet afterglow parameters), \ref{fig:light_curve_corner_plot_KN} (kilonova parameters) and \ref{fig:light_curve_corner_plot_full} (all parameters).

\begin{figure}
    \centering
    \includegraphics[width=\hsize]{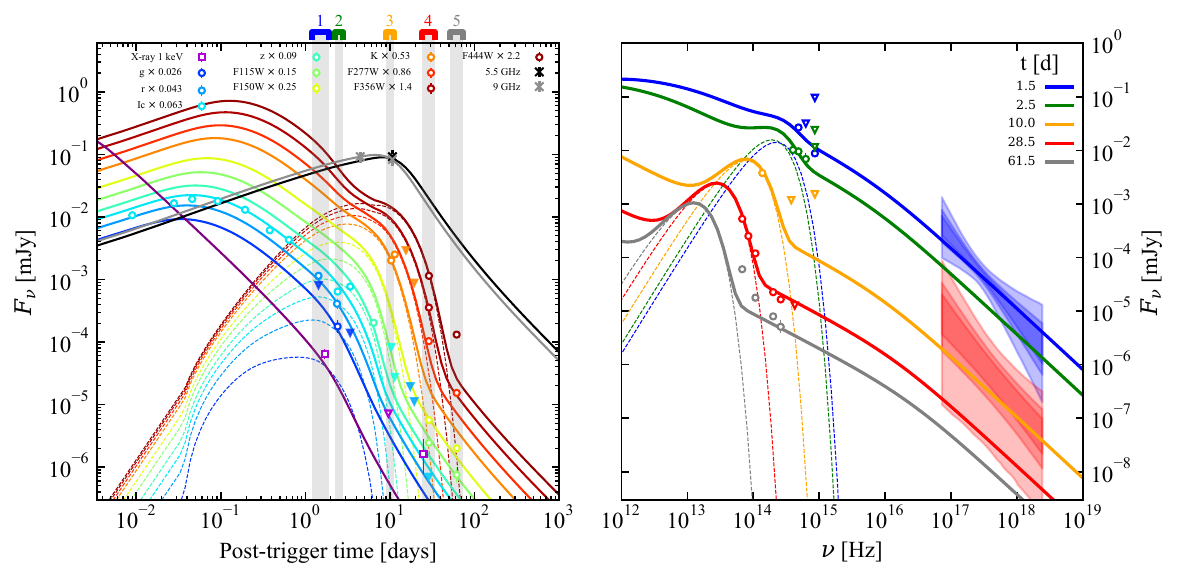}
    \caption{Multi-wavelength light curves and model predictions. Markers in the figure show the observed flux density at the position of GRB~230307A in various bands (see legend in the left-hand panel) and at various times. Downward-facing triangles represent upper limits. The optical and near infrared flux densities are multiplied by the numbers reported in the legend for presentation purposes. The butterfly-shaped filled regions in the right-hand panel encompass flux densities consistent at one, two and three sigma (progressively lighter shades) with the {\em Swift}/XRT and Chandra detections in the 0.3-10 keV band, according to our analysis and adopting a uniform prior on the flux. Solid lines of the corresponding colours show the predicted light curves (left-hand panel) and spectra (right-hand panel) of our afterglow (forward shock only) plus kilonova model at the central frequencies of the bands. Dashed lines single out the contribution of the kilonova.}
    \label{fig:light_curve_model}
\end{figure}

The left-hand panel in Figure \ref{fig:light_curve_model} shows the observed light curve data (markers) along with the best-fitting model (solid lines). Dashed lines single out the contribution of the kilonova. The right-hand panel in the same figure shows some selected spectra, showing in particular the good agreement of the first JWST epoch with the blackbody plus power law spectrum implied by our model at those times. 

While the best-fit model demonstrates a relatively good agreement with most of the measurements, some discrepancies stand out, most prominently with the 61.5 d JWST data and with the 28.5 d \emph{Chandra} detection. The former is not too surprising, as the assumptions in the kilonova model (in particular that of blackbody photopsheric emission, which is particularly rough in such a nebular phase, and that of constant and uniform grey opacity, due to recombination of at least some species) are expected to break down at such late epochs. The latter is linked to the steepening (`jet break') apparent at around 2 days in the model X-ray light curve, which in turn is mainly driven by the need to not exceed the optical and near-infrared fluxes implied by observations at around one week and beyond. In absence of these constraints, the fit would have accommodated a larger jet half-opening angle, postponing the jet break and hence allowing for a better match with the best-fit \textit{Chandra} flux. On the other hand, as noted in Methods, this flux is rather uncertain, with the low-end uncertainty possibly extending to fluxes lower by one order of magnitude or more, depending on the adopted prior in the spectral analysis (see Methods). Still, such a discrepancy might indicate the presence of additional X-ray emission that is not accounted for by the model, as has been seen previously in e.g.\ \cite{perley09,fong14}.

\subsection{Spectral analysis modeling}\label{sec:spec_model}
The JWST/NIRSpec  spectrum taken on 5 April 2023 exhibits a red continuum component with emission line features. The most distinctive feature is a broad emission line at
$2.15$ microns (in the rest frame, assuming $z=0.065$). This may be 
a blend (visibly split in Figure\,\ref{NIRSpec}) and a simultaneous fit of two Gaussians provides measured 
centroids of 20285$\pm10$\AA\ and 22062$\pm10$\AA.  The line widths are both consistent 
at  $v_{\rm FWHM}=19100$\,kms$^{-1}$ (0.064c).  This $2.1$\,micron feature is 
quite similar in strength and width to the $2.07$\,micron feature  in AT2017gfo at 10.5 days after merger \citep[discussed in][]{Gillanders23}. The AT2017gfo line 
also appears to be better
fit as a blend of two features rather than a single transition, with line velocities of 
$v_{\rm FWHM}=38900$\,kms$^{-1}$. While the average line centre is reasonably 
consistent between the two, the components inferred for AT2017gfo and the kilonova of
GRB 230307A are each quite different. Reference \cite{Gillanders23} finds them 
at 20590\AA\ and  21350\AA\ and there is no consistent velocity shift that could be 
applied to match AT2017gfo with our JWST spectrum. 
Nevertheless, the similarity in their average 
line centroids, velocities and equivalent widths  is
striking, as demonstrated in Figure\,\ref{NIRSpec}.

With a Doppler broadening parameter of $\lesssim 0.1c$,  it is unlikely that the continuum component is formed as a result of the superposition of emission lines. 
Because kilonova radiation transfer at such late times is not yet fully understood, here we attempt to model the spectrum with the assumption that the emission  consists of blackbody radiation from the photosphere and forbidden emission lines of heavy elements formed  outside the photosphere. 

If the continuum is described with blackbody radiation, the temperature and photospheric velocity are $\approx 670\,{\rm K}$ and $\approx 0.08c$, respectively. The continuum luminosity is estimated as $\sim 2\times 10^{39}\,{\rm erg/s}$ in the NIRSpec band and $\sim 5\times 10^{39}\,{\rm erg/s}$ if the blackbody emission extends to much longer wavelengths.  Assuming this emission is entirely powered by radioactivity of $r$-process nuclei, these correspond to an ejecta mass of $\sim 0.03$--$0.07M_{\odot}$ \cite{hotokezaka20}. With the ejecta mass and velocity, the opacity is required to be $\gtrsim 5\,{\rm cm^2/g}$ in order to keep the ejecta optically thick at $30$ day. It is worth noting that such a high opacity in the mid-IR indicates that the inner part of the ejecta is lanthanide rich \cite{kasen13,tanaka13,fontes20}.

\begin{figure}
   \centering
   \includegraphics[width=\hsize]{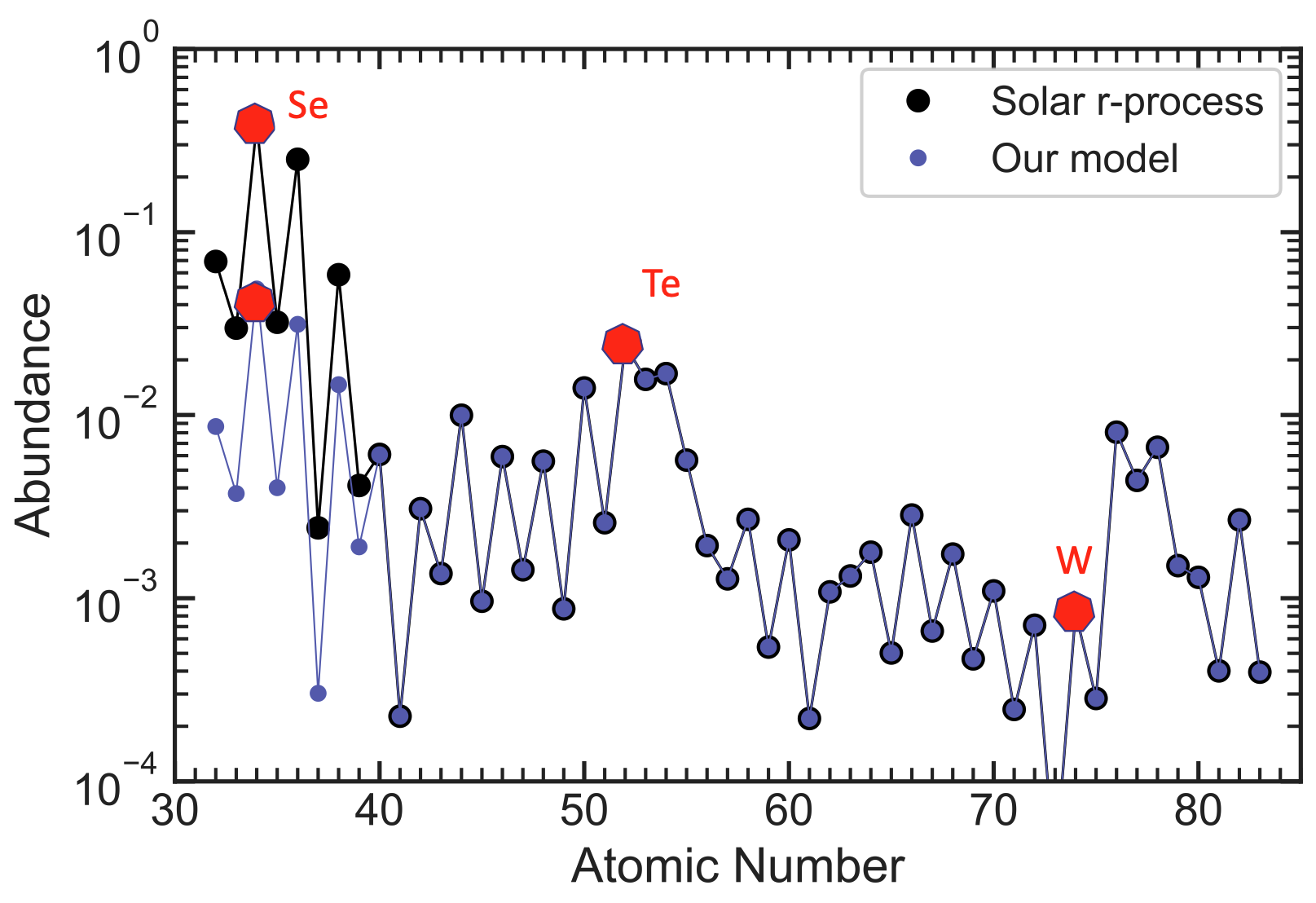}
      \caption{Abundance used in the spectral modeling. The abundance is chosen based on the solar r-process residuals. The abundance of the ``light'' elements ($A<85$) is reduced relative to the solar pattern. The locations of selenium (Se), tellurium (Te) and tungsten (W) are marked.  }\label{fig:abundance}
   \end{figure}

Forbidden emission lines in the infrared are expected to arise from fine structure transitions of low-lying energy levels of heavy elements. Most abundant ions are expected to produce the strongest lines. We attribute the strongest observed line at 2.15 microns to tellurium (Te) III from an M1 line list of heavy elements presented in \cite{hotokezaka22}, where the line wavelengths are experimentally calibrated according to the NIST database \cite{NIST_ASD}. Te  belongs to the second $r$-process peak.
With the M1 line list, we model kilonova emission line spectra under the assumption that photons from forbidden lines produced outside the photosphere freely escape from the ejecta. The collision strengths of Te III are taken from an R-matrix calculation \cite{Madonna2018} and those of other ions are obtained by using an atomic structure code HULLAC\cite{hullac}. The abundance pattern is chosen to be the solar $r$-process but we separate ``light'' and ``heavy'' elements at an atomic mass of $85$ and introduce a parameter, the abundance ratio of the two (see figure \ref{fig:abundance}). 
The ionization fractions are fixed to be $(Y^{+1},Y^{+2},Y^{+3})=(0.2,0.5,0.3)$ motivated by the Te ionization evolution in  kilonova ejecta \cite{Pognan2022MNRAS}. The line shape is approximated by a Gaussian with a line broadening velocity of $0.08c$, which is the same as the photospheric velocity. 
The mass in the line forming region is estimated by assuming that the observed line luminosity, $5\times 10^{38}\,{\rm erg\,s^{-1}}$, is locally generated by radioactivity of $r$-process nuclei, corresponding to $\sim 0.02M_{\odot}$. Given the abundance pattern and ionization state, the mass of Te III in the line forming region is $\approx 8\cdot 10^{-4}M_{\odot}$.
The electron temperature of the line forming region is then determined such that the total line luminosity agrees with the observed one. The estimated electron temperature is $\sim  3000\,{\rm K}$, which is slightly higher than that derived from the pure neodymium nebular modeling \cite{hotokezaka21}. This is because the cooling by tellurium ions is more efficient than neodymium. 

We find that [Te III] $2.10\,{\rm \mu m}$ line is indeed the most outstanding emission line around 2 microns. 
Several weaker lines also contribute to the flux around $3$--$4$ microns. There is another potential line feature around $4.5$ microns in the NIRSpec spectrum. The location of this feature is consistent with  [Se III] $4.55\,{\rm \mu m}$ and [W III] $4.43\,{\rm \mu m}$ as pointed out by \cite{hotokezaka22} for the kilonova AT2017gfo. From the spectral modeling, we obtain the total ejecta mass of $\sim 0.05$--$0.1M_{\odot}$, which agrees with the one obtained from the light curve modeling $\sim 0.1M_{\odot}$.

Here we show a brief estimate of the Te III mass from the observed line at $2.15$ microns (${\rm ^3P_0}$--${\rm ^3P_1}$). The collisional excitation rate per Te III ion from the ground level (${\rm ^3P_0}$) to the first exited level (${\rm ^3P_1}$) is given by
\begin{align}
    k_{01} = \frac{8.63\cdot 10^{-6}n_e}{\sqrt{T_e}}\frac{\Omega_{01}}{g_{0}} e^{-E_{01}/kT_e}\,{\rm s^{-1}},
\end{align}
where $n_e$ and $T_e$ are the thermal electron density and temperature, $\Omega_{01}\approx 5.8$ is the collision strength \cite{Madonna2018}, $E_{01}\approx 0.6\,{\rm eV}$ is the excitation energy, $g_0$ is the statistical weight of the ground level. Assuming that the ejecta mass in the line forming region is $0.02M_{\odot}$ expanding with $0.08c$ and the ions are typically doubly ionised, we estimate $n_e\sim 3\cdot 10^{5}\,{\rm cm^{-3}}$, and thus, the line emissivity per Te III ion is
\begin{align}
    \epsilon_{10} \approx 2.5\cdot 10^{-14}\left(\frac{n_e}{3\cdot 10^5\,{\rm cm^{-3}}}\right)\,{\rm erg/s},
\end{align}
where $T_e=3000\,{\rm K}$ is used. Combining the line emissivity with the observed line luminosity in $2.25\pm 0.23\,{\rm \mu m}$, $L_{\rm line}\approx 3\cdot 10^{38}\,{\rm erg/s}$, we obtain
\begin{align}
    M({\rm Te\,III}) \approx 10^{-3}M_{\odot}
    \left(\frac{n_e}{3\cdot 10^5\,{\rm cm^{-3}}}\right)^{-1}
    \left(\frac{L_{\rm line}}{3\cdot 10^{38}\,{\rm erg/s}}\right).
\end{align}
The mass estimated from the line is somewhat dependent on $T_e$ and $n_e$. However, we emphasise that, with $T_e\approx3000\,{\rm K}$ and $n_e\approx 3\cdot 10^{5} {\rm cm^{-3}}$, the line luminosity is consistent with the radioactive power in the line forming region.
It is also interesting to note that the Te III mass of $10^{-3}M_{\odot}$ is in good agreement with the one obtained based on the same line seen in AT2017gfo at 10.5 day \cite{hotokezaka23}.

While we conclude that the observed line feature at 2.1 microns is most likely attributed to Te III, it is important to note that there are caveats associated with our modeling. One obvious caveat is that the model does not include E1 lines. 
Lanthanides and actinides have E1 transitions between low-lying levels in the mid-IR
\citep[discussed in][]{Gillanders23}. Due to their lower abundances, these lines are expected to be weaker compared to the Te line if collisional excitation dominates the excitation processes.  
However, as we make an implicit assumption that their E1 lines   contribute to the opacity in the mid-IR, they may produce P-Cygni like features, see, e.g., \cite{domoto22,Gillanders23}. 
For example, Ce III has a strong line at $2.07\,{\rm \mu m}$ with $\log gf=-1.67$ \cite{domoto22}. We estimate that its line optical depth is $\lesssim 0.1$ at $700\,{\rm K}$ with $\sim 0.05M_{\odot}$ and $\sim 0.1c$ even if Ce is purely in Ce III. However, more careful analyses including non-LTE effects are needed to quantify it.
Another caveat is that the  opacity of lanthanides is expected to have some wavelength dependence. Including this effect may also  affect the spectral modelings.


\section{Alternative progenitor possibilities}
Our interpretation of GRB 230307A provides a self-consistent model for the source in which the temporal and spectral evolution, as well as the source location, can be readily explained. The kilonova has marked similarities with AT2017gfo providing a robust indication of its origin, and we do not need to postulate new and unseen phenomena to explain it. However, it is also relevant to consider alternative possibilities. In particular, given the location of the galaxy at $z=3.87$, it is important to consider if the burst could originate at that redshift. 

\subsection{GRB 230307A as a high redshift, highly energetic GRB}
\label{hiz}
The nearby galaxy H1 (F277W(AB)=27.5$\pm$ 0.1 , $r_{proj} = 0.3$ arcsec) with a spectroscopic redshift of z=3.87 has a relatively low probability of occurring by chance ($\sim 5-10$\%, see section~\ref{host_con}). This galaxy has a comparable P$_{chance}$ to G1 (the $z=0.065$ galaxy). The host-normalised offset for H1 is $\sim 2.5$, which is large but not unprecedented for long GRBs \citep{blanchard16}. However, assuming the late time light at the GRB position is all from the transient, it does not lie on the stellar field of this galaxy, which is unusual, for example, in the samples of  \citep{fruchter06,blanchard16,lyman17}, there is only one (of $>100$) sub-arcsecond localised GRB not on the stellar field of its host.

At $z=3.87$, the inferred isotropic energy release and luminosity of GRB 230307A would be $E_\mathrm{iso} = 
1.2 \times 10^{56}$ erg and $L_\mathrm{iso} = 1.7 \times 10^{56}$ erg s$^{-1}$ (using a 64 ms peak flux). This is approximately an order of magnitude more energetic and two orders of magnitude more luminous than any other previously identified GRB \citep{burns23}. 

If at $z=3.87$, we can have some confidence that GRB 230307A would be the most energetic burst ever detected by {\em Swift} or {\em Fermi}, including those without redshift or even afterglow identifications. In Table~\ref{tab:bright_bursts}, we tabulate the most fluent GRBs observed by {\em Fermi}. Most of these have either redshifts or optical detections, which constrain $z<6$ via the detection of the source in the optical band. This leads to a set of measured or maximum $E_\mathrm{iso}$ values. For events without any redshift information, we can place a conservative upper redshift limit of $z = 16$. No GRBs detected by {\em Fermi} without a redshift can have energy over $10^{56}$ erg unless they lie beyond $z \sim 20$. Hence, GRB 230307A is sufficiently rare, if at $z=3.87$, that events like it occur less frequently than once per decade across the Universe (i.e. no more than one in the combined lifetimes of {\em Swift} and {\em Fermi}). 

The energetics of the burst at this redshift lie would lie far beyond those of the general GRB population and beyond those suggested as the upper limit for GRB energetics \cite{atteia17}. The only population of core-collapse GRBs whose energetics have been suggested to approach this value are those from first-generation population III stars \cite{desouza11,toma11,toma16}. It is not expected that such stars should exist at $z \sim 4$. However, while a pop-III origin may alleviate energy concerns, the properties of the GRB and its optical/IR counterpart do not resemble the predictions for pop-III stars. In particular, pop III GRBs are suggested to have particularly long durations given the mass and radii of their progenitors \cite{toma16}, and so require extremely long durations of engine activity to enable jet breakout. However, the $\sim$ 35 s duration of GRB 230307A and its rapid variability do not readily fit this expectation. 

If one ascribes the GRB to a stellar collapse event, considering the afterglow's properties and associated supernovae is also relevant. Firstly, at 28.5 days, the JWST spectral observations are inconsistent synchrotron emission, suggesting that the counterpart must be dominated by another source in the mid-IR and the earlier K-band points. This excess, which in our preferred model is explained by a kilonova, would have to be due to the supernova or shock breakout if at $z=3.87$. The K-band (rest frame B-band) light would reach a peak of $M_B(AB) < -23.5$ on a timeframe of $<$2-days (rest-frame) before decaying at a rate of $>$ 1 mag day$^{-1}$ for the next four days, or as a power-law decay, a rate of approximately $t^{-4}$. This appears too rapid for radioactively powered transients, at least based on the sample to date (note that at $z=3.87$, the timescales are a factor of $\sim 5$ faster than in the $z=0.065$ scenario due to cosmological time dilation). The most likely option for such emission would be shock-breakout, which may begin blue but rapidly cool. There are simulations for the shock breakout associated with pop-III supernovae, which show an early peak \cite{tolstov16}. However, this emission peaks in the UV to soft X-ray regime and at luminosities below that seen in GRB 230307A. Indeed, taking 28.5 days as a baseline; for a plausible maximum Pop III radius (e.g. 2000 $R_{\odot}$, \cite{ohkubo09}) and a luminosity of $L_{bol} \sim 10^{43}$ erg s$^{-1}$ the inferred temperature is $T \sim 300,000$\, K. This is incompatible with the spectral shape seen in the counterpart to GRB 230307A which peaks at $>4.5$ microns ($T< 3000$K at $z=3.87$). Dust or metal line blanketing could alleviate this discrepancy to some degree, but it would be extreme to explain the observations. It would also come at the cost of an even higher intrinsic luminosity. Conversely, the radius at which the luminosity and temperature would be consistent is extremely large ($\sim 0.3$ pc) and indeed would require super-luminal expansion to reach from a single explosion within the time since burst. These constraints become even more extreme for the K-band observations at 11.5 days, where the luminosity is $>50$ times higher. However, we lack detailed information regarding the spectral shape at this time. We can conclude that a thermal transient launched at the time of GRB 230307A cannot explain the observed source at $ z=3.87$.  


We should consider if GRB 230307A could be related to an explosion which bears little to no similarity to long-GRB progenitors. Given the inferred energetics, this is not an unreasonable proposition. However, the emission is too bright and too fast for, for example, the fast blue optical transients (e.g. \cite{perley18}), the fastest of which have half-times of $\sim 4$ days \citep[][c.f. $<1$ day for GRB 230307A at $z=3.87$]{ho23}. A further alternative may be a relativistic tidal disruption event. This would face significant challenges with the rapid variability timescales seen in the prompt emission and the non-nuclearity of the source within the galaxy at $z=3.87$. Putting aside these concerns, the peak optical/IR luminosity is comparable to AT2022cmc \citep{andreoni23,pasham23}, but the evolution is too rapid and the dynamic range too large. 

Finally, it is possible that the red excess seen at later times is not directly related to the progenitor or the transient but is a result of the re-processing of the GRB radiation by material within the host galaxy. In particular, for GRB 211211A \cite{waxman23} suggest that an alternative explanation for the emission could be the heating of dust. However, this model also encounters significant issues at $z = 3.87$. In particular, the observed $K$-band excess is a rest-frame B-band excess, much bluer than expected for dust heating. If this represented the peak of the thermal spectrum, it would be above the sublimation temperature of the dust. Alternatively, if it were the blue tail of a much cooler black body, the luminosity would be extremely high. 

A final challenge to the high$-z$ scenario is that the afterglow is detected in the UVOT-white and ground-based $g$-bands. These observations all have substantial sensitivity blue-ward of Ly$\alpha$ at z=3.87. A typical column from the intergalactic medium should attenuate $\sim 50\%$ of the light in these bands, inconsistent with observations. Indeed, for a typical $\beta = 1$ spectrum, we expect to observe $white-i$ = 2.8 and $g-i=1.9$, approximately 3 and 4 $\sigma$ away from the observed colours. There is significant variation in the absorption strength as a function of the line of sight, so a low (or near zero) absorption column would alleviate this tension. The sample of \cite{thomas20} implies that at $z \sim 4$, perhaps 10\%  of galaxies have such low absorption sight lines. 

Hence, while the proximate galaxy could indicate a high redshift, there are few other indications in the transient properties that would support this interpretation. In particular, neither standard thermal nor non-thermal emission can explain the observed counterpart properties. If the burst does arise from $z=3.87$, it requires a new kind of explosion, unlike any seen until now. In practice, such explosions could be extremely rare: the volumetric rate of GRBs with $E_\mathrm{iso} > 10^{56}$ erg is minimal, but postulating them is unnecessary when a robust, physically motivated explanation can be obtained for a lower redshift solution.

\subsection{Other cosmological scenarios}
We should also consider a further option which is that GRB 230307A does not reside at either $z=0.065$ or $z=3.87$ but is a chance super-position with both galaxies. In this scenario, the actual host is undetected or is one of the other galaxies within the field. The absence of direct redshift measurements makes placing constraints on this scenario challenging. However, we can use the non-detection of the late-time JWST magnitudes to limit the brightness of a supernova component at any redshift.

\begin{figure}
   \centering
   \includegraphics[width=\hsize]{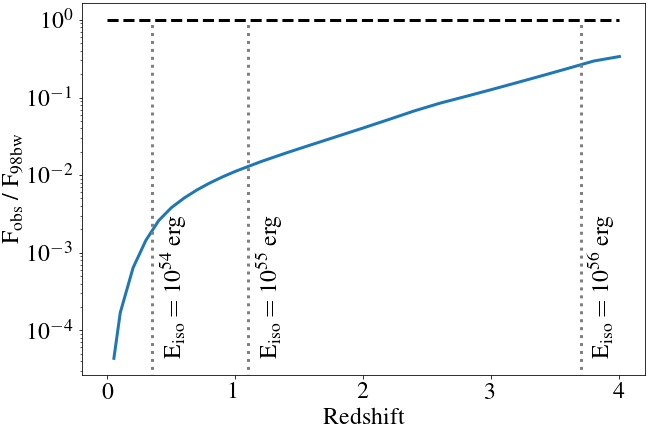}
      \caption{Limits on supernova similar to SN~1998bw as a function of redshift, based on the most constraining detection with {\em JWST}. At any redshift for which GRB 230307A would not be the most energetic GRB ever observed, any supernova is at least a factor of $\sim$ 100 fainter than SN~1998bw.}\label{fig:98bw_constraints}
   \end{figure}

To quantify the exclusion of ``normal" long GRBs at intermediate redshift we utilize model light curves for SN 1998bw from MOSFiT \cite{nicholl17} calculated at a range of redshift from $0.05 < z < 4.0$ (we take $z =4$ as an upper limit for the redshift the GRB based on the observed g-band detections together with the detection of continuum emission to 5300\AA\ \cite{gillandersGCN} and similar faint trace seen to $\sim$5100\AA in our X-shooter spectrum). At each redshift, we compare our observed JWST photometry in each band to the model predictions at that time and report the most constraining limit (e.g. the lowest ratio of $F_{obs} / F_{98bw}$). This is shown in Figure~\ref{fig:98bw_constraints}. At all plausible redshifts, any supernova must be at least a factor $\sim 3$ fainter than SN~1998bw at similar times. For any redshift where the burst energetics fall within the range seen in the bulk GRB population ($z<1.2$), any supernova must be a factor $> 100$ fainter than SN~1998bw. Hence, there is no route in which GRB 230307A can arise from a classical long GRB ($E_\mathrm{iso} < 10^{55}$ erg with an associated broad-lined Ic supernova. The strength of this rejection is predominantly based on the faintness of the source in the bluer bands, whereas at the first epoch, the redder bands are substantially brighter than SN~1998bw at $z=3.87$. 

If the burst lies at an intermediate redshift, dust models may become more appealing. In particular, for a low to moderate redshift, the luminosty and timescales may be a suitable match (e.g. for GRB 211211A \cite{waxman23} find plausible explanations at $z \sim 0.5$). However, in this case, the 
lack of a supernova to extremely deep limits would be surprising, as would the non-detection of the host galaxy. 

\subsection{Galactic objects}

In the absence of a robust absorption redshift, it is also necessary to consider if GRB 230307A could arise from a Galactic system. 

The very faint magnitudes and extreme red colours observed at late times can effectively rule out X-ray binary outbursts. For example, with an M-dwarf companion (absolute magnitude 9), the distance to the source would be $\sim 100$ kpc, and larger for any more massive star, while the late-time colours of the source are not stellar. In practice, given that the source is transient, and we may also expect some contribution from an accretion disc, the overall properties cannot be remedied within the accreting binary framework. 

The inferred energetics for Galactic systems are $E_\mathrm{iso} \sim 5 \times 10^{43} (d/10\mathrm{\, kpc})^2$. This energetic output is within the bounds of giant outbursts from magnetars. For example, the giant flare of SGR 1806-20 had an inferred isotropic energy of $E_\mathrm{iso} \sim 2 \times 10^{46}$ erg (see \cite{palmer05}, although subsequent downward revisions in its distance lower this somewhat \cite{bibby08}). However, GRB 230307A does not appear to meet the requirements of a magnetar. In particular, it is at high Galactic latitude (-36 degrees) and far from any plausible star formation that could give rise to a young neutron star. Furthermore, the emission in all magnetar outbursts is dominated by a very short pulse followed by decaying emission in which the pulse period of the neutron star is visible. This is not the case for GRB 230307A. 

GRBs have previously been suggested to arise from the tidal disruption and accretion of rocky material onto a neutron star \cite{campana11}, and such events are seen in the case of white dwarfs \cite{gaensicke19}. However, for accretion onto a neutron star, we would usually expect to observe a relatively soft outburst (e.g. in the model of \cite{campana11} the temperature is $\sim 10$ keV). The spectrum we observe for GRB 230307A consists of evolving synchrotron emission which does not contain a thermal component and is inconsistent with directly observing heating accreting material. Indeed, even for near direct accretion, it is not clear how such a spectrum would be formed in an accreting neutron star scenario. Indeed, in this case, the evolution to very low temperatures on a timescale of $\sim 60$ days would also not be natural. 

Hence, we conclude that no known Galactic systems could explain the observed properties of GRB 230307A. 

\subsection{GRB 230307A as a white-dwarf -- neutron star merger}
A final alternative is that GRB 230307A is related to the merger of a white dwarf and a neutron star. Although this is still a ``compact object merger", such mergers are very different from those of neutron stars with another neutron star or a black hole. In particular, simulations show that no $r$-process material is produced \cite{2022arXiv220913061K}, and so we should not expect the very red emission. Although there are suggestions that GRB 211211A could have been produced by a WD-NS merger \cite{yang22,zhong23}, it is unclear if these could readily explain the detailed spectrophotometric evolution of GRB 230307A. 

White dwarf neutron star mergers are appealing, because the long-duration of the gamma-ray emission could suggest a less compact remnant merger event.  The wider separation of the binary at the disruption of the white dwarf produces disk accretions times from 100 to 1000s, matching the long duration of these bursts~\cite{1999ApJ...520..650F}.  However, we expect the accretion rate in the mergers to be low, producing less-powerful, and hence, less luminous GRBs.  Current WD/NS merger simulations predict a range of light curves that span the emission from GRB\,230307A.  A Ca feature does 
exist that is close to the observed line feature,  
but the models are too blue to explain the shape of the spectra~\cite{2022arXiv220913061K}.  This is because WD/NS mergers do not produce elements much heavier than the iron peak elements.  As we have noticed in matching kilonova models, these elements do not have strong lines beyond $\sim  20{,}000$~\AA. The subsequent emission above these wavelengths is very weak in these models.  In addition, WD/NS mergers are expected to have fairly week kicks to ensure that the binary remains bound and these mergers are expected to have much lower offsets than neutron star mergers.  However, the mass of the best-candidate host galaxy is sufficiently low that the observed offset for this burst can be attained~\cite{2018A&A...619A..53T}.

    \begin{table*}
    \centering
    \begin{tabular}{cccccc}
    \hline\hline
    Time since GRB & Telescope & Band & Exposure time & Magnitude & Source \\
    (days) & & & (s) & (AB) & \\
    \hline
    0.01 & TESS & $I_{\rm C}$ & 1600 & $18.63 \pm 0.14$ & This work \\
    0.03 & TESS & $I_{\rm C}$ & 1600 & $18.15 \pm 0.12$ & This work \\
    0.04 & TESS & $I_{\rm C}$ & 1600 & $17.98 \pm 0.11$ & This work \\
    0.09 & TESS & $I_{\rm C}$ & 6400 & $18.06 \pm 0.10$ & This work \\
    0.19 & TESS & $I_{\rm C}$ & 11200 & $18.41 \pm 0.10$ & This work \\
    0.38 & TESS & $I_{\rm C}$ & 20800 & $19.23 \pm 0.11$ & This work \\
    0.64 & TESS & $I_{\rm C}$ & 24000 & $19.61 \pm 0.16$ & This work \\
    0.99 & UVOT & $U$ & 2315$^{\dagger}$ & $>21.1$  & This work \\
    1.16 & UVOT & WH & 692$^{\dagger}$ & $22.25^{+0.23}_{-0.19}$ & This work \\
    1.25 & UVOT & WH & 2217$^{\dagger}$ & $22.29^{+0.34}_{-0.26}$ & This work \\
    1.43 & ULTRACAM     & $u$ & 200  & $>19.7$ & This work \\
    1.43 & ULTRACAM     & $g$ & 200  & $>20.7$ & This work \\
    1.43 & ULTRACAM     & $r$ & 200  & $20.72 \pm 0.15$ & This work \\
    1.61 & UVOT & WH & 6468$^{\dagger}$ &$ >22.60$ & This work  \\
    2.37 & VST          & $r$ & 360  & $21.84 \pm 0.19$ & This work \\
    2.41 & ULTRACAM     & $u$ & 200  & $>21.2$ & This work \\
    2.41 & ULTRACAM     & $g$ & 200  & $22.35 \pm 0.26$ & This work \\
    2.41 & ULTRACAM     & $i$ & 200  & $21.68 \pm 0.09$ & This work \\
    2.43 & UVOT & WH & 3303$^{\dagger}$ & $>22.0$  & This work \\
    3.39 & ULTRACAM     & $u$ & 200  & $>20.8$ & This work \\
    3.39 & ULTRACAM     & $g$ & 200  & $>22.6$ & This work \\
    3.39 & ULTRACAM     & $i$ & 200  & $21.48 \pm 0.18$ & This work \\
    4.89 & UVOT & WH & 14921$^{\dagger}$ & $>23.6$ & This work  \\
    6.42 & FORS2        & $z$ & 1440 & $23.24 \pm 0.11$ & This work \\
    6.42 & FORS2        & $B$ & 1380 & $>26.10$  & This work \\
    10.34 & FLAMINGOS-2 & $K$ &  840 & $22.51 \pm 0.15$ & This work \\
    10.36 & FORS2       & $I$ & 2400 & $>24.0$ & This work \\
    11.36 & FORS2       & $I$ & 2400 & $>25.2$ & This work \\
    11.42 & FLAMINGOS-2 & $K$ & 700 & $22.27 \pm 0.15$ & This work \\
    15.45 & FLAMINGOS-2 & $K$ & 950 & $> 22.1$ & This work \\
    17.38 & FORS2 & $R$ & 3000 & $>25.2$ & This work\\
    19.37 & FORS2 & $R$ & 3000 & $>25.8$ & This work\\
  19.38 & HAWK-I & $K$ & 2340 & $>23.4$ & This work \\
  \hline
    28.89   & JWST  & F070W & 1868 &  28.97 $\pm$ 0.20  & This work  \\
    28.83    & JWST & F115W & 1868 & 28.50 $\pm$ 0.07   & This work \\
    28.86    & JWST & F150W  &1546 &  28.11 $\pm$ 0.12  & This work \\
    28.83   & JWST & F277W & 1868 & 26.24 $\pm$ 0.01  & This work\\
    28.86   & JWST & F356W & 1546 & 25.42 $\pm$ 0.01  & This work\\
    28.89    & JWST & F444W & 1868 & 24.62 $\pm$ 0.01  & This work \\   

     61.48    & JWST & F115W & 1868 & 29.78 $\pm$ 0.31 & This work \\
    61.51   & JWST & F150W  &1546 &  29.24 $\pm$ 0.17  & This work \\
   61.51   & JWST & F277W & 1868 & 28.31 $\pm$ 0.12  & This work\\
    61.48   & JWST & F444W & 1546 & 26.97 $\pm$ 0.04  & This work\\
    
    \hline
    2.35 & GMOS-S       & $r$ & 30   & $22.0 \pm 0.3$ & \cite{2023GCN.33447....1O} \\
    2.35 & SOAR         & $z$ & 310  & $21.8 \pm 0.3$ & \cite{2023GCN.33459....1B} \\
    3.35 & SOAR         & $z$ & 600  & $21.8 \pm 0.3$ & \cite{2023GCN.33459....1B} \\
    \hline\hline
    \end{tabular}
    \caption{Optical and IR observations of the optical counterpart of GRB 230307A. Limits are given at the $3\sigma$ level.}
    \label{tab:photometry}
\end{table*}

\begin{table*}
    \centering
    \begin{tabular}{ccccc}
    \hline\hline
    Time since trigger & Telescope & Frequency & Flux density & Source \\
    (days) & & (Hz) & ($\mu$Jy) & \\
    \hline \\
    1.7  & {\em Swift}/XRT & $2.42\times10^{17}$ & $(6.5\pm 1.1)\times 10^{-2}$ & This work \\
    9.59 & {\em Swift}/XRT & $2.42\times10^{17}$ & $<$ $7.14 \times 10^{-3}$ & \cite{2023GCN.33465....B} \\
    24.84 & {\em Chandra} & $2.42\times10^{17}$ &   1.6$^{+1.2}_{-0.8} \times 10^{-3}$ & This work \\

    4.46 & ATCA & $5.5\times10^9$ & $< 84$ & This work \\
    4.46 & ATCA & $9\times10^9$ & $92 \pm 22$ & This work \\
    10.66 & ATCA & $16.7\times10^9$ & $< 114$ & This work \\
    10.66 & ATCA & $21.2\times10^9$ & $< 165$ & This work \\
    10.69 & ATCA & $5.5\times10^9$ & $92 \pm 36$ & This work \\
    10.69 & ATCA & $9\times10^9$ & $83 \pm 26$ & This work \\
    25.55 & ATCA & $16.7\times10^9$ & $< 81$ & This work \\
    25.55 & ATCA & $21.2\times10^9$ & $< 219$ & This work \\
    25.59 & ATCA & $5.5\times10^9$ & $< 63$ & This work \\
    25.59 & ATCA & $9\times10^9$ & $< 63$ & This work \\

    6.64 & MeerKAT & $1.3\times10^9$ & $< 390$ & This work \\
    6.75 & MeerKAT & $3.1\times10^9$ & $< 140$ & This work \\
    15.74 & MeerKAT & $1.3\times10^9$ & $< 350$ & This work \\
    16.04 & MeerKAT & $3.1\times10^9$ & $< 120$ & This work \\
    40.89 & MeerKAT & $3.1\times10^9$ & $< 93$ & This work \\
    
    \hline\hline
    \end{tabular}
    \caption{X-ray and radio observations of the afterglow of GRB 230307A.}
    \label{tab:radio_xrays}
\end{table*}
\begin{table*}
    \centering
    \begin{tabular}{ccccccc}
    \hline\hline
    Host candidate & RA & Dec & $r$ & $z$ & offset ($^{\prime \prime}$) & $P_{chance}$ \\
    \hline
    H1 & 04:03:26.06 & -75:22:42.5 & $>29$ & 3.87 & 0.30 & 0.05*\\
    G1 & 04:03:18.79 & -75:22:55.0 & 17.6 & 0.0645 & 29.9 & 0.09\\
    G2 & 04:03:27.32 & -75:23.09.3 & 18.6 & 0.2633 & 27.0 & 0.15\\
    G3 & 04:03:25.64 & -75:23:17.0 & 18.8 & 0.2626 & 34.2 & 0.27\\
    G4 & 04:03:16.67 & -75:22:23.2 & 19.4 &  0.2627 & 29.9 & 0.32  \\
    \hline
    \end{tabular}
    \label{tab:hosts}
    \caption{Properties of possible host galaxies for GRB 230307A. *Formally, because the galaxy is undetected in the $r-$band, $P_{chance}$ is unbounded. This probability is based on the magnitudes measured at other wavelengths.}
\end{table*}

\begin{table*}
    \centering
    \begin{tabular}{ccccccc}
    \hline\hline

    Burst  & fluence$\times 10^{-3}$ & t$_{90}$ & $F_X$ &$z$ & $E_\mathrm{iso}$  & Notes \& Refs \\
    & (erg cm$^{-2}$) & (s) & (11 hour) &  &  ($10^{52}$ erg) & \\
    \hline
    221009A & 21 & 325.8 & 1.1 $\times 10^{-9}$ & 0.15 & 1150 & SN ambiguous (1,2,3)  \\
    230307A &  3.2  & 34.6 & 6.0 $\times 10^{-13}$ & 0.065/3.87 &2.34/12000 & This work\\
    130427A & 2.46 & 138.2 & 1.0 $\times 10^{-10}$ & 0.34   &72.3 & Spectroscopic SN (4,5) \\
    160625A & 0.64 & 43.0 & 1.1 $\times 10^{-12}$ &1.41 & 336.2 & Too high-z for SN/KN (6)\\
    171010A & 0.63 & 107.3 & 1.5 $\times 10^{-11}$ & 0.33 & 17.4 & Spectroscopic SN likely (7) \\
    160821A & 0.52 & 43.0 & - & $<6$ & $<1300$ & No SN/KN\\
    211211A & 0.50 & 34.3 & 4.7 $\times 10^{-12}$ &0.08 & 0.74 &  Kilonova (8,9,10) \\
    190114C & 0.43 & 116.4 &1.1 $\times 10^{-11}$ & 0.42 & 19.6 &  Spectroscopic SN  (11) \\ 
    190530A & 0.37 & 18.43 & 2.5 $\times 10^{-11}$  & $<4.5$ &  $<1400$ & No afterglow\\
    221023A & 0.34 & 39.2 & - & $<16$ & $<7800$ & No afterglow  \\
    \hline
    \end{tabular}
    \caption{The properties of the brightest 10 bursts observed by 
    {\em Fermi}/GBM. Eight of these have afterglow identifications which place limits on the redshift, while it is likely that all originate from $z<16$. As expected for the (observationally) brightest bursts there is a preference for low redshift, with the bursts all arising at much less than the mean for {\em Swift} GRBs 
    \citep{jakobsson06}. In 3-4 cases a supernova has been identified, meaning the bursts arise from core collapse (with GRB 221009A still ambiguous at the time of writing). In 2 cases a kilonova origin appears more likely, while for the remaining 4 the absence of afterglows or the redshifts render such diagnostics impossible. However, it would appear that samples of bright, long-GRBs may contain significant number of compact object mergers.}
    \label{tab:bright_bursts}
\end{table*}

\begin{table*}
    \centering
    \begin{tabular}{ccc}
    \hline\hline
    Telescope/filter & Magnitude & Uncertainty  \\
    \hline
    GALEX/$FUV$  & 20.87 & 0.15 \\
    GALEX/$NUV$  & 20.38 & 0.13 \\
    UVOT/$u$     & 20.09 & 0.30 \\
    LS/$g$       & 18.28 & 0.02 \\
    LS/$r$       & 17.80 & 0.03 \\
    LS/$z$       & 17.34 & 0.05 \\
    JWST/$F070W$ & 17.98 & 0.01 \\
    JWST/$F115W$ & 17.49 & 0.01 \\
    JWST/$F150W$ & 17.32 & 0.01 \\
    JWST/$F277W$ & 17.75 & 0.01 \\
    JWST/$F356W$ & 17.93 & 0.01 \\
    JWST/$F444W$ & 18.24 & 0.01 \\
    WISE/$W1$    & 17.86 & 0.05 \\
    WISE/$W2$    & 18.23 & 0.12 \\
    \hline
    \end{tabular}
    \label{tab:hostmag}
    \caption{Photometry (in the AB photometric system) of the host galaxy of GRB 230307A. No reddening correction was applied.}  
\end{table*}

\begin{table*}
    \centering
    \begin{tabular}{ccc}
    \hline\hline 
        Parameter & Prior bounds & Median \& 90\% C.\,I.  \\
        \hline
        $\log(E_\mathrm{K}/\mathrm{erg})$ &  $(49.5,55)$ & $50.7_{-1.2}^{+1.0}$  \\
        $\log(n/\mathrm{cm^{-3}})$ & (-6,2) & $-0.6^{+2.0}_{-2.0}$ \\
        $\theta_\mathrm{j}/\mathrm{rad}$ & (0.01,0.5) & $0.23^{+0.12}_{-0.14}$ \\
        $\log(\Gamma_0)$ & $(2,4)$ & $3\pm 1$ \\
        $\log(\xi_\mathrm{N})$ & $(-2,0)$ & $-1.5_{-0.5}^{+0.8}$ \\
        $\log(\epsilon_\mathrm{e})$ &  $(-4,-0.5)$ & $-1.2_{-2.2}^{+0.6}$ \\
        $\log(\epsilon_\mathrm{B})$ & $(-6,-0.5)$ & $-2.2_{-2.0}^{+1.4}$ \\
        $p$ & $(2.01,2.99)$ & $2.39^{+0.42}_{-0.25}$ \\
        $\log(M_\mathrm{ej}/\mathrm{M_\odot})$ & $(-4,0)$ & $-1.23^{+0.57}_{-1.4}$ \\
        $\log(v_\mathrm{ej}/c)$ & $(-3,-0.3)$ & $-1.0^{+0.51}_{-0.74}$ \\
        $\log(\kappa/\mathrm{cm^2\,g^{-1}})$ & $(-1,2)$ & $-0.17^{+1.1}_{-0.73}$ \\
        $\delta$ & $(1.5,7)$ & $4.4_{-2.6}^{+2.4}$ \\
        $\log(f_\mathrm{sys})$ & $(-4,0)$ & $-0.25^{+0.12}_{-0.09}$\\  
        \hline
    \end{tabular}
    \caption{Light curve model parameters, priors and posterior medians and 90\% credible intervals.}
    \label{tab:light_curve_mcmc}
\end{table*}

\clearpage

\section*{Author contribution statements}
AJL led the project, including the location of the afterglow and kilonova and the JWST observations. BPG first identified the source as a likely compact object merger, was co-PI of the \emph{Chandra} observations, and contributed to analysis and writing. OS contributed to afterglow and kilonova modelling and led the writing of these sections. MB was involved in kilonova modelling, EB contributed to interpretation, placing the burst in context and high energy properties. KH was involved in kilonova spectral modelling and identified the 2.15 micron feature. LI reduced VLT/MUSE and X-shooter observations and led the host analysis. GPL contributed to afterglow and kilonova modelling. ARE analysed the Chandra observations, BS reduced and analysed VLT observations. NS contributed to afterglow and kilonova modelling. SS was responsible for placing the burst afterglow in context and demonstrating its faintness. NRT contributed to analysis, interpretation and writing. KA was involved in the ULTRACAM observations and interpretation. GA led the ATCA observations. GB reduced the JWST NIRCAM data. LC processed and analysed the MUSE observations. VSD is the ULTRACAM PI. JPUF contributed to the interpretation. WF was the PI on the Chandra observations and contributed to discussion. CF contributed to the theoretical interpretation. NG was involved in host analysis. KEH, GP, AR, SDV, SC, PDA, DH, MDP, CCT, AdUP and DA contributed to ESO observations and discussion. DW contributed to spectral and progenitor modelling and discussion. MJD, PK, SP, JM, SGP, IP, DIS contributed to the ULTRACAM observations.
GL investigated potential similarities with other transients. AT, PAE, BS and JAK contributed to the {\em Swift} observations. 
MF extracted and flux-calibrated the TESS light curve. SJS analysed the JWST spectral
lines, and contributed to interpretation and writing. 
HFS performed the BPASS-hoki-ppxf fits to the integrated MUSE flux and contributed the associate figure and text. 
All authors contributed to manuscript preparation through contributions to concept development, discussion and text. 

\section*{Acknowledgements}
We dedicate this paper to David Alexander Kann, who passed on March 10. The final messages he sent were regarding follow-up of GRB 230307A, and we hope it would satisfy his curiosity to know the final conclusions.  

This work is based on observations made with the NASA/ESA/CSA James Webb Space Telescope. The data were obtained from the Mikulski Archive for Space Telescopes at the Space Telescope Science Institute, which is operated by the Association of Universities for Research in Astronomy, Inc., under NASA contract NAS 5-03127 for JWST. These observations are associated with program \#4434 and 4445. Support for Program numbers 4434 and 4445 was provided through grants from the STScI under NASA contract NAS5- 03127.

This paper is partly based on observations collected at the European Southern Observatory under ESO programme 110.24CF (PI Tanvir), and on observations obtained at the international Gemini Observatory (program IDs GS-2023A-DD-106), a program of NOIRLab, which is managed by the Association of Universities for Research in Astronomy (AURA) under a cooperative agreement with the National Science Foundation on behalf of the Gemini Observatory partnership: the National Science Foundation (United States), National Research Council (Canada), Agencia Nacional de Investigaci\'{o}n y Desarrollo (Chile), Ministerio de Ciencia, Tecnolog\'{i}a e Innovaci\'{o}n (Argentina), Minist\'{e}rio da Ci\^{e}ncia, Tecnologia, Inova\c{c}\~{o}es e Comunica\c{c}\~{o}es (Brazil), and Korea Astronomy and Space Science Institute (Republic of Korea). Processed using the Gemini \texttt{IRAF} package and \texttt{DRAGONS} (Data Reduction for Astronomy from Gemini Observatory North and South). 

AJL, DBM and NRT were supported by the European Research Council (ERC) under the European Union’s Horizon 2020 research and innovation programme (grant agreement No.~725246).

N.~Sarin is supported by a Nordita fellowship. Nordita is supported in part by NordForsk.
B.~Metzger is supported in part by the NSF (grant AST-2002577).
J.H. and D.L. were supported by a VILLUM FONDEN Investigator grant (project number 16599).
G.P.L. is supported by a Royal Society Dorothy Hodgkin Fellowship (grant Nos. DHF-R1-221175 and DHF-ERE-221005).
G.L. was supported by a research grant (19054) from VILLUM FONDEN.
K. H. is supported by JST FOREST Program (JPMJFR2136) and the JSPS Grant-in-Aid for Scientific Research (20H05639, 20H00158, 23H01169, 20K14513).
KEH acknowledges support from the Carlsberg Foundation Reintegration Fellowship Grant CF21-0103.
S. Schulze acknowledges support from the G.R.E.A.T. research environment, funded by {\em Vetenskapsr\aa det},  the Swedish Research Council, project number 2016-06012.
The Cosmic Dawn Center (DAWN) is funded by the Danish National Research Foundation under grant No. 140. JPUF is supported by the Independent Research Fund Denmark (DFF–4090-00079) and thanks the Carlsberg Foundation for support.
SJS acknowledges funding from STFC Grant ST/X006506/1 and ST/T000198/1.
VSD and ULTRACAM are funded by STFC grant ST/V000853/1
AAB acknowledges funding from the UK Space Agency.
MN is supported by the European Research Council (ERC) under the European Union’s Horizon 2020 research and innovation programme (grant agreement No.~948381) and by UK Space Agency Grant No.~ST/Y000692/1.
H.F.S is supported by the Eric and Wendy Schmidt AI in Science Postdoctoral Fellowship, a Schmidt Futures program.
DS acknowledges funding from STFC grants ST/T000406/1, ST/T003103/1, ST/X001121/1.
MER acknowledges support from the research programme Athena with project number 184.034.002, which is financed by the Dutch Research Council (NWO). POB acknowledges funding from STFC grant ST/W000857/1.
D.K.G acknowledges support from the Australian Research Council Centre of Excellence for Gravitational Wave Discovery (OzGrav), through project number CE170100004.

\section*{Data availability}
JWST data are directly available from the MAST archive. {\em Chandra} and {\em Swift} data are also in the public domain. ESO and Gemini data are stored in their respective archives and will be available to all once the proprietary period expires. Data can be obtained from the corresponding author between the date of publication and the end of the proprietary period. 
This research has made use of \emph{Fermi} data which are publicly available and can be obtained through the High Energy Astrophysics Science Archive Research Center (HEASARC) website at \url{https://heasarc.gsfc.nasa.gov/W3Browse/fermi/fermigbrst.html}

\section*{Code availability}
Much analysis for this paper has been undertaken with publically available codes and the details required to reproduce the analysis are contained within the manuscript.

\clearpage 

\bibliography{refs}

\end{document}